\documentclass[12pt,english]{article}
\usepackage[T1]{fontenc}
\usepackage[latin1]{inputenc}
\usepackage{geometry}
\usepackage{babel}
\usepackage{subfigure}
\usepackage{graphics}
\usepackage{rotating}
\usepackage{amssymb}

\def\be{\begin{equation}}
\def\beq{\begin{equation}}
\def\eeq{\end{equation}}
\def\ee{\end{equation}}

\def\R{{\bf R}}
\def\beqn{\begin{eqnarray}}
\def\eeqn{\end{eqnarray}}
\def\ba{\begin{eqnarray}}
\def\ea{\end{eqnarray}}

\def\slash#1{#1\hskip-6pt/\hskip6pt}

\def\one {\bf  1}

\newcommand{\beqa}{\begin{eqnarray}}
\newcommand{\eeqa}{\end{eqnarray}}

\makeatletter

\providecommand{\LyX}{L\kern-.1667em\lower.25em\hbox{Y}\kern-.125emX\@}

\makeatother
\begin{document}
\begin{center}
\vspace{3.cm}
{\bf \Large Parton Distributions, Logarithmic Expansions and Kinetic Evolution  \\}
\vspace{2cm}
 {\large \bf $^{1}$Alessandro Cafarella, $^{2}$Claudio Corian\`{o} and $^{2}$Marco Guzzi\\}
\vspace{1.cm}
{\it $^1$Department of Physics, University of Crete, 71003 Heraklion, Greece\\}
\vspace{0.5cm}
{\it $^2$Dipartimento di Fisica, Universit\`{a} di Lecce and INFN sezione
di Lecce\\
Via per Arnesano, 73100 Lecce, Italy\\}
\vspace{0.5cm}
{\it cafarell@physics.uoc.gr, claudio.coriano@le.infn.it, marco.guzzi@le.infn.it\\}
\end{center}

\vspace{2.cm}
\begin{abstract}
Aspects of the QCD parton densities are briefly reviewed, drawing some parallels 
to the density matrix formulation of quantum mechanics, exemplified by Wigner functions. 
We elaborate on the solution of their evolution equations using logarithmic expansions 
and overview their kinetic interpretation.
We illustrate how a Fokker-Planck equation can be derived using the master
formulation of the same equations and its construction in the case of the transverse spin distributions. A simple connection of the leading order 
DGLAP equation to fractional diffusion using fractional calculus is also 
briefly outlined. 
\footnote{To appear in {\bf Lecture Notes of SIM, Seminario Interdisciplinare di Matematica}, 
S. Dragomir (ed.)}

\end{abstract}
\newpage
\section{Introduction}

In Quantum Chromodynamics, or QCD, the accepted theory of the strong interactions,
the description of high energy collisions is formulated with the help of some important
theorems, commonly refered to as {\em factorization theorems} \cite{Sterman}. According to these theorems 
it is possible to describe certain hadronic processes in three steps,  
the first involving the initial state, followed by a second intermediate state, commonly known as ``the hard scattering'', 
and the third which involves the final state. Both the initial and the final state are described with the help 
of some non-local operators supported on the light cone, whose matrix elements, 
termed ``parton distributions'' and ``fragmentation functions'' respectively, 
can't be computed from first principle, at least at this time,  
in a second quantized theory. It is possible, however, to parametrize their functional form 
by combining a vast amount of experimental data with numerical fits. 

The separation between the initial and the intermediate state is perfomed with the 
introduction of an intermediate scale, the factorization scale, denoted by $Q$, where $Q$ 
is the amount of energy/virtuality involved in the actual separation. 

Parton distributions
depend both on the choice of $Q$ and on a second variable, $x$ (Bjorken x) which is supported on the interval $(0,1)$. This variable can be interpreted as the fraction of the momentum 
of the particles in the initial state that flows into the intermediate state.
Parton distributions, at least some of them, have a simple interpretation in terms of 
densities of quarks and gluons (the partons) in a nucleon. They are probability distributions that change 
(evolve) with the scale $Q$ and their evolution is described by a set of equations known 
as DGLAP equations \footnote{Dokshitzer-Gribov-Lipatov-Altarelli-Parisi}, which are a special type of renormalization group equation (RGE). They are, by now, textbook material in high energy physics. Mathematicians may find a thorough 
discussion of their origin in \cite{Sterman} and in other books in Quantum Field Theory and particle phenomenology. 

On the 
contrary of other RGE's, such as those describing the evolution of other local operators in the 
Standard Model which are ordinary differential equations, 
the RGE's of the parton densities are integro-differential equations. They are 
defined in terms of some kernels $(P(x))$ which are not regular functions but 
distributions, finite after integration over $x$.

The objective of this work is to present a short review of 
previous studies by us on this subject. Here we will discuss first a direct way to solve these equations and then move 
to describe their kinetic interpretation. We will show how one 
can construct a hierarchy of equations starting from the original integro-differential 
equations and illustrate the procedure at the lowest order and identify a 
Fokker-Planck approximation to the same equations. A simple connection between the leading order equations and the 
formalism of fractional calculus is also briefly pointed out.

\section{Parton distributions as Wigner functions}
Evolution equations describing the high energy behaviour 
of scattering amplitudes carry significant information on the factorization/ renormalization scale  
dependence of such amplitudes, and allow to link the behaviour 
of processes at a given energy scale to collisions taking place at another (usually much higher) scale.

In QCD confinement forbids the detection of the fundamental 
states of the theory, such as quarks and gluons. However, 
asymptotic freedom, the smallness of the coupling constant as the energy raises (or approaching smaller distances),  allows to separate the perturbative 
dynamics at short distances from the non-perturbative one, due to confinement, 
through factorization theorems and the introduction of the parton distributions. 
We will very briefly introduce them below, 
and we will exploit the density matrix formulation of quantum mechanics 
as an analogy to illustrate the topic. 

The mathematical construct which is the closest 
to a parton distribution function (p.d.f.) $q(x, Q)$ 
is a {\it Wigner function}, as first noticed in \cite{Gallipoli}. The analogy is, of course, limited.  

We recall that Wigner's description of quantum mechanics 
via quasi-probabilities of phase space $(\bf{x},p)$
\beq
f(\bf{x},p)= \frac{1}{2 \pi}\int dy \psi^*(x - \frac {\not{h}}{2}y)
e^{i p\cdot y}\psi(x + \frac {\not{h}}{2}y )
\label{wwg}
\eeq
is fully equivalent to Schrodinger's formulation \cite{zachos}.
Its dynamics is specified by Moyal's equation, that extends Liouville's formulation of classical mechanics for a classical hamiltonian $H({\bf x},{\bf p})$, 
$\partial_t f + \{f,H\}=0$ to quantum mechanics 
\beq
\frac{\partial f}{\partial t}=\frac{H*f - f*H}{i \slash{h}}
\label{dyn}
\eeq
where the star-product (*) is 
\beq
*\equiv e^{i \frac{\slash{h}}{2}\left(\overleftarrow{\partial}_x\overrightarrow
{ \partial_p} - \overleftarrow{\partial}_p\overrightarrow{ \partial}_x\right)}.
\eeq
The nonlocality of the construct appears immediately in its dynamics, 
with derivatives in its evolution equation (\ref{dyn}) that extend to 
all orders. In practice one can use translation of its arguments  
\beq
f(x,p)*g(x,p)=f\left(x + i \frac{\slash{h}}{2} \overrightarrow{\partial}_p, p -  i \frac{\slash{h}}{2}\overrightarrow{\partial}_x\right)
g(x,p)
\eeq
for the explicit evaluation of this product. 

Differently from Wigner functions, in a p.d.f. the variable $x$ takes now the role of the momentum ``p''.
Also, parton distributions are correlation functions of a special type, 
being defined just on the light-cone. In this sense they are not generic 
nonlocal correlators. This limitation, due to the special nature of 
high energy collisions in asymptotically free theories, sets the boundary  
of validity of the parton model approach to QCD. 

Another observation that may help in the distinction 
between a p.d.f. and a Wigner function is the different nature of the respective equations. For the p.d.f. 
the nonlocality in the evolution is all expressed in a renormalization group equation rather than in the Hamiltonian 
formalism of (\ref{dyn}). However, also in this case some interesting analogies remain. Once the evolution equation of a 
p.d.f. is written in the form of a master equation, a hierarchy of equations can be identified quite straighforwardly 
using the Kramers-Moyal expansion of the transition probabilities. This point will be discussed next. 
From this analysis, that we believe has limited 
phenomenological applications, emerge however some amusing features which may be generic for other types of nonlocal operators in field theory.   
\subsection{ The ``Wiggies'' of QCD}

In QCD one starts by introducing, via arguments based on unitarity, the hadronic tensor, 
which is the key construct describing the collision

\beq
W^{\mu\nu}=\int {d^4 x\over 2 (\pi)^4}e^{i q\cdot x} 
\langle P_A S_A;P_B S_B|\left[J^\mu(0),J^{\nu}(x)\right]| P_A S_A;P_B S_B
\rangle,
\eeq

with $P_A$ and $P_B$ being the momenta of the colliding hadrons and 
$S_A$ and $S_B$ their covariant spins. The $J$'s  are electromagnetic currents.

The distribution functions that emerge -at leading order- 
from this factorized picture are 
correlation functions of non-local operators in configuration space. 
They are the quark-quark and the quark-antiquark 
correlators.

Their expression simplifies in the axial gauge, in which the 
gauge link is removed by the gauge condition. For instance, the quark-quark correlator takes 
the form 

\beq
\left(\Phi_{a/A}\right)_{\alpha\beta}(P,S,k)=\int {d^4z\over (2 \pi)^4} 
e^{i k\cdot z}
\langle P,S|\overline{\psi}^{(a)}_\beta (0)\psi^{(a)}_\alpha(z)|PS\rangle.
\label{bu}
\eeq
In (\ref{bu}) we have included the quark flavour index $a$ and an index $A$ for the hadron, 
as usual. 
Fields are not time ordered since they can be described by the good 
light cone components of $\psi$ and by $A_T$, a transverse component of the 
gauge field $A_\mu$, as discussed in \cite{RLJ}. 
The non-perturbative information in a collision is carried by 
matrix elements of this type. 

Further considerations allow to show that 
the leading contributions to (\ref{bu}) come from the light-cone region.
The leading expansion of the quark-quark correlator then is of the form

\beq
\int {d\lambda\over 2 \pi}e^{i\lambda x}\langle PS|
\overline{\psi}(0)\psi(\lambda n)|P S\rangle={1\over 2}
\left( \not{p}f_1(x) +\lambda \gamma_5 \not{p} g_1(x) +
\gamma_5 \not{S_T}\not{p}h_1(x)\right), \nonumber \\
\nonumber\\
\label{pdf}
\eeq
where we have used all the four vectors at our disposal 
(spin S, momentum P of the hadron) and introduced invariant amplitudes 
(parton distributions) 
$f_1$, $g_1$, $h_1$, now expressed in terms of a scaling variable x 
(Bjorken x). $n^\mu$ is a light-cone four-vector $(n^2=0)$, approximately 
 perpendicular to the hadron momentum.

The definition of p.d.f.'s in (\ref{pdf}) involves also an underlying physical scale $Q$ 
($Q >> \Lambda_{QCD}$, with $\Lambda_{QCD}$ being the scale of confinement), not apparent from that equation and characterizing the energy scale 
at which these matrix elements, summarized by (\ref{pdf}), are defined. 
Truly: $f_1=f_1(x,Q^2)$, $h_1=h_1(x,Q^2)$ and so on. 

The role of the 
various renormalization group equations 
is to describe the perturbative change in 
these functions as the scale $Q$ is raised (lowered). Each equation 
involves kernels $(P(x))$ of various types, of well known form, and 
asymptotic expansions of the solutions exist 
(see for instance \cite{FP} and the implementation given in \cite{CS}). 
The structure to all orders of the solution  has been discussed by us 
recently using alternative methods \cite{Cacorguz}.

\subsection{Factorization and Evolution} 
Factorization theorems play a crucial role in the application of perturbation theory to hadronic 
reactions. The proof of these theorems and the actual implementation of their implications has spanned a 
long time and has allowed to put the parton model under a stringent experimental test. Prior to embark on 
the discussion of our contributions to the study of evolution algorithms in Bjorken $x$-space, 
we provide here a brief background on the topic in order to make our treatment self-contained.
We refer to \cite{Sterman} for a thorough overview of the topic.

In sufficiently inclusive cross section, leading power factorization theorems allow to write down a 
hadronic cross section in terms of parton distributions and of some hard scatterings, the latter being 
calculable at a given order in perturbation theory using the fundamental QCD lagrangean. 
Specifically, for a hadronic cross section, for instance  a proton proton cross section $\sigma_{pp}$,
the result of the calculation can be summarized by the formula 
\ba
\sigma_{pp} =\sum_{f} \int^{1}_{0} dx_{1} \int^{1}_{0} dx_{2} f_{h_{1}\rightarrow f}(x_{1},Q^{2}) f_{h_{2}\rightarrow f}(x_{2},Q^{2})\hat{\sigma }(x_{1},x_{2},\hat{s},\hat{t},Q^2)\,,
\label{pr}
\ea
where the integral is a function of some variables $x_1$ and $x_2$ which describe the QCD dynamics at parton level in the Deep Inelastic Scattering (DIS) limit or, equivalently, at large energy and momentum transfers. 
These variables are termed ``Bjorken variables'' and are scale invariant. 
 This formula is a statement about the computability of a hadronic collision in terms of 
some ``building blocks'' of easier definition.

The variable $Q^2$, in the equation above, can be identified with the factorization scale of the process. $\hat{\sigma}$ can be computed at a given order in perturbation theory in an expansion  in $\alpha_s$, the QCD coupling, while the $f_{h_i\to f}(x, Q^2)$ are the parton distributions. These describe the probability for a hadron $h$ to prepare for the scattering
a parton $f$, which undergoes the collision. An equivalent interpretation of the functions 
$f_{h_i\to f}(x, Q^2)$ is to characterize the density of partons of type $f$ into a hadron of type $h$. 
A familiar notation, which simplifies the previous notations shown above, 
is to denote by $q_i(x,Q^2)$ the density of quarks in a hadron (a proton in this case) 
of flavour $i$ and by $g(x,Q^2)$ the corresponding density of gluons. For instance, the annihilation 
channel of a quark with an antiquark in a generic process is accounted for by the contribution 
\be
\int_0^1 dx_1 \int_0^2 dx_2 q(x_1,Q^2) \overline{q}(x_2,Q^2) \hat{\sigma}_{q \overline{q}}(x_1,x_2,Q^2),
\ee
 and so on for the other contributions, such as the quark-gluon sector (qg) or the gluon-gluon sector 
(gg) each of them characterized by a specific hard scattering cross section $\hat{\sigma}_{qg}$,  or 
$\hat{\sigma}_{qg}$. In this separation of the cross section into contributions of hard scatterings 
$\hat{\sigma}$ and parton distributions $f(x,Q^2)$ the scale at which the separation occurs 
is artificial, in the sense that the hadronic cross section $\sigma$ should not depend on $Q^2$ or 
\be
\frac{d \sigma}{d Q^2} =0.
 \ee
However, a perturbative computation performed by using the factorization formula, however, shows that this 
is not the case, since the perturbative expansion of $\hat{\sigma}$
\be
\hat{\sigma}= \hat{\sigma}^{(0)} +\alpha_s(Q^2)^2 \hat{\sigma}^{(1)}+ \alpha_s(Q^2)\hat{\sigma}^{(2)}\qquad
\ee 
naturally brings in the dependence on the factorization scale Q. This dependence is weaker if we are able 
to push our computation of the hard scattering to a sufficiently high order in $\alpha_s$. The order at which the perturbative expansion stops is also classified as a ``leading order'' (LO), 
``next-to-leading order'' (NLO) or, even better, a ``next-to-next-to-leading'' (NNLO) 
contribution if more and more terms in the expansion are included. 

At the same time, as we have already mentioned, 
the parton distributions $f(x,Q^2)$ also carry a similar dependence on the 
scale $Q$, which is summarized by some RGE's. 
The equations resum the logarithmic violations 
to the lowest order scale invariance, induced by the perturbative expansion.
Also in this case we need to quantify this effect and reduce its impact on the prediction of the cross section. Solving the RGE's for the p.d.f.'s to higher order and, at the same time, computing the hard scatterings to higher orders reduces the spurious dependence on the factorization scale $Q$ and improves the theoretical prediction of the real physical phenomenon.

\subsection{Parton dynamics at NLO}

As we have mentioned, 
within the framework of the parton model, where the ``partons'' are the quarks and gluons inside a hadron, evolution equations of DGLAP-type
- and the corresponding initial conditions on the parton distributions -
are among the most important
blocks which characterize the description of
the quark-gluon interaction.
Other parts of this description require the computation of the hard scatterings
(what we have called ``the second stage'' of the collision) 
with high accuracy. 
Here we illustrate an implementation to NLO of a method based on an
ansatz which allows to rewrite the evolution
equations as a set of recursion relations for some scale invariant
functions, \( A_{n}(x) \) and \( B_{n}(x) \), which appear in the
expansion. The advantage, compared to others, 
of using these recursion relations is that just few iterates 
of these are necessary in order to obtain a stable solution. 
One of the advantages of the method is its analytical base,  
since the recursion relations can be solved explicitely in terms of the initial conditions. 
We have recently shown conclusively that these methods are equivalent to other methods that also 
give the exact solution using Mellin transforms. Our approach follows closely the Mellin method 
in the demonstration of the correctness of the recursion relations, but then is implemented directly 
in x-space. 

\section{Running coupling}
In QCD the coupling constant changes with the change of the energy scale and there is a RGE associated to it as well, computed within a certain level of accuracy, also classified as LO, NLO, NNLO etc. 
The solution of the RGE for the running coupling constant in the NLO approximation can be expressed 
in terms of two coefficients $\beta_0$ and $\beta_1$, describing the beta function of the theory and a 
renormalization group invariant scale $\Lambda_{QCD}$ that sets the scale for confinement. 
The NNLO extension of the equation involves a third coefficient, $\beta_2$. The RGE of 
the coupling is 

\begin{equation}
\beta(\alpha_{s})=\frac{\textrm{d}\alpha_{s}(Q^{2})}{\textrm{d}\log Q^{2}},\label{eq:beta_def}\end{equation}
and its expansion, for instance to NNLO, is given by
\cite{betafunction} \cite{alpha_s}
\begin{equation}
\beta(\alpha_{s})=-\frac{\beta_{0}}{4\pi}\alpha_{s}^{2}-\frac{\beta_{1}}{16\pi^{2}}\alpha_{s}^{3}-\frac{\beta_{2}}{64\pi^{3}}\alpha_{s}^{4}+O(\alpha_{s}^{5}),\label{eq:beta_exp}.\end{equation} For the moment we work in the NLO 
approximation (only $\beta_0$ and $\beta_1$ are included), and in this case an explicit solution 
of this equation is given by
\begin{equation}
\label{eq:alpha_s}
\alpha _{s}(Q^{2})=\frac{4\pi }{\beta _{0}}\frac{1}{\log (Q^{2}/\Lambda _{\overline{MS}}^{2})}\left[ 1-\frac{\beta _{1}}{\beta ^{2}_{0}}\frac{\log \log (Q^{2}/\Lambda _{\overline{MS}}^{2})}{\log (Q^{2}/\Lambda _{\overline{MS}}^{2})}+O\left( \frac{1}{\log ^{2}(Q^{2}/\Lambda _{\overline{MS}}^{2})}\right) \right] ,
\end{equation}
where\begin{equation}
\beta _{0}=\frac{11}{3}N_{C}-\frac{4}{3}T_{f},\qquad \beta _{1}=\frac{34}{3}N^{2}_{C}-\frac{10}{3}N_{C}n_{f}-2C_{F}n_{f},
\end{equation}
and\begin{equation}
N_{C}=3,\qquad C_{F}=\frac{N_{C}^{2}-1}{2N_{C}}=\frac{4}{3},\qquad T_{f}=T_{R}n_{f}=\frac{1}{2}n_{f},
\end{equation}
where \( N_{C} \) is the number of colors, \( n_{f} \) is the number
of active flavors, which is fixed by the number of quarks with \( m_{q}\leq Q \).
$\Lambda_{QCD}$ is a scale that depends on the number of active quarks included in the evolution 
(number of flavours).
We have taken for the quark masses (charm, bottom and top 
respectively) $ m_{c}=1.5\, \textrm{GeV}, m_{b}=4.5\, \textrm{GeV}$ 
and $ m_{t}=175\, \textrm{GeV} $, these are necessary in order 
to identify the thresholds at which the number of flavours $n_f$ 
is raised as we increase the final evolution scale.

$\Lambda_{QCD}$ should be more correctly denoted as $\Lambda _{\overline{MS}}^{(n_{f})}$, given 
its flavour dependence, and is given by

\begin{equation}
\Lambda _{\overline{MS}}^{(3,4,5,6)}=0.248,\, 0.200,\, 0.131,\, 0.050\, \textrm{GeV}.
\end{equation}
The label ``$\overline{MS}$'' denotes conventionally the modified minimal subtraction scheme 
and denotes a specific regularization scheme of the perturbative expansion. 

We also define the distribution of a given helicity $(\pm)$,
\( f^{\pm }(x,Q^{2}) \), which is the probability 
of finding a parton of type \( f \) at a scale \( Q \), 
where \( f=q_{i},\overline{q_{i}},g \),
in a longitudinally polarized proton with the spin aligned (+) 
or anti-aligned (-) respect 
to the proton spin and carrying a fraction \( x \) of the proton's
momentum. 

We introduce the longitudinally polarized parton
distribution of the proton\begin{equation}
\label{eq:long_distr}
\Delta f(x,Q^{2})\equiv f^{+}(x,Q^{2})-f^{-}(x,Q^{2}).
\end{equation}
We also introduce another type of parton density, termed 
{\em transverse spin 
distribution}, which is defined as 
the probability of finding a parton of type \( f \) in a transversely
polarized proton with its spin parallel ($\uparrow$) minus the probability 
of finding it antiparallel ($\downarrow$) to the proton
spin
\begin{equation}
\Delta _{T}f(x,Q^{2})\equiv f^{\uparrow }(x,Q^{2})-f^{\downarrow }(x,Q^{2}).
\end{equation}

Similarly, the unpolarized (spin averaged) parton distribution of the proton
is given by
\begin{equation}
\label{eq:unp_distr}
f(x,Q^{2})\equiv f^{+}(x,Q^{2})+f^{-}(x,Q^{2})=f^{\uparrow }(x,Q^{2})+f^{\downarrow }(x,Q^{2}).
\end{equation}
We also recall, if not obvious, that 
taking linear combinations of Equations (\ref{eq:unp_distr}) and
(\ref{eq:long_distr}), one recovers 
the parton distributions of a given helicity 
\begin{equation}
f^{\pm }(x,Q^{2})=\frac{f(x,Q^{2})\pm \Delta f(x,Q^{2})}{2}.
\end{equation}
In regard to the kernels, the notations \( P \), \( \Delta P \), \( \Delta _{T}P \),
\( P^{+\pm} \),  will be used to denote the 
kernels in the unpolarized, longitudinally polarized, transversely
polarized, and the positive (negative) helicity cases respectively.

The DGLAP equation is an integro-differential equation whose general
mathematical structure is

\begin{equation}
\frac{\textrm{d}}{\textrm{d}\log Q^{2}}f(x,Q^{2})=P(x,\alpha_{s}(Q^{2}))\otimes f(x,Q^{2}),\label{eq:DGLAP}\end{equation}
where the convolution product is defined by\begin{equation}
\left[a\otimes b\right](x)=\int_{x}^{1}\frac{\textrm{d}y}{y}a\left(\frac{x}{y}\right)b(y)=\int_{x}^{1}\frac{\textrm{d}y}{y}a(y)b\left(\frac{x}{y}\right).\end{equation}

Let us now turn to the evolution equations, starting from the unpolarized
case. Defining\begin{equation}
\label{eq:definizioni}
q_{i}^{(\pm )}=q_{i}\pm \overline{q_{i}},\qquad q^{(+)}=\sum _{i=1}^{n_{f}}q_{i}^{(+)},\qquad \chi _{i}=q_{i}^{(+)}-\frac{1}{n_{f}}q^{(+)},
\end{equation}
the evolution equations are  
\begin{equation}
\frac{\textrm{d}}{\textrm{d}\log Q^{2}}q_{i}^{(-)}(x,Q^{2})=P_{NS^{-}}(x,\alpha _{s}(Q^{2}))\otimes q_{i}^{(-)}(x,Q^{2}),
\end{equation}
\begin{equation}
\frac{\textrm{d}}{\textrm{d}\log Q^{2}}\chi _{i}(x,Q^{2})=P_{NS^{+}}(x,\alpha _{s}(Q^{2}))\otimes \chi _{i}(x,Q^{2}),
\end{equation}
for the non-singlet sector and
\begin{equation}
\label{eq:singlet}
\frac{\textrm{d}}{\textrm{d}\log Q^{2}}\left( \begin{array}{c}
q^{(+)}(x,Q^{2})\\
g(x,Q^{2})
\end{array}\right) =\left( \begin{array}{cc}
P_{qq}(x,\alpha _{s}(Q^{2})) & P_{qg}(x,\alpha _{s}(Q^{2}))\\
P_{gq}(x,\alpha _{s}(Q^{2})) & P_{gg}(x,\alpha _{s}(Q^{2}))
\end{array}\right) \otimes \left( \begin{array}{c}
q^{(+)}(x,Q^{2})\\
g(x,Q^{2})
\end{array}\right) 
\end{equation}
for the singlet sector.
Equations analogous to (\ref{eq:definizioni}-\ref{eq:singlet}),
with just a change of notation, are valid in the longitudinally polarized
case and, due to the linearity of the evolution equations, also for
the distributions in the helicity basis. In the transverse case instead,
there is no coupling between gluons and quarks, so the singlet sector
(\ref{eq:singlet}) is missing. In this case we will solve just the
nonsinglet equations \begin{equation}
\frac{\textrm{d}}{\textrm{d}\log Q^{2}}\Delta _{T}q_{i}^{(-)}(x,Q^{2})=\Delta _{T}P_{NS^{-}}(x,\alpha _{s}(Q^{2}))\otimes \Delta _{T}q_{i}^{(-)}(x,Q^{2}),
\end{equation}
\begin{equation}
\frac{\textrm{d}}{\textrm{d}\log Q^{2}}\Delta _{T}q_{i}^{(+)}(x,Q^{2})=\Delta _{T}P_{NS^{+}}(x,\alpha _{s}(Q^{2}))\otimes \Delta _{T}q_{i}^{(+)}(x,Q^{2}).
\end{equation}

We also recall that the perturbative expansion, up to next-to-leading order, of the kernels
is\begin{equation}
P(x,\alpha _{s})=\left( \frac{\alpha _{s}}{2\pi }\right) P^{(0)}(x)+\left( \frac{\alpha _{s}}{2\pi }\right) ^{2}P^{(1)}(x)+\ldots .
\end{equation}
Kernels of fixed helicity can also be introduced
with $P_{++}(z)=(P(z)+\Delta P(z))/2$ and $ P_{+-}(z)=(P(z)-\Delta P(z))/2$ 
denoting splitting functions of fixed helicity, which will be used below.

 The equations, in the helicity basis, are 

\begin{eqnarray}
{dq_+(x) \over{dt}}=
{\alpha_s \over {2 \pi}} (P_{++}^{qq} ({x \over y}) \otimes q_+(y)+
P_{+-}^{qq} ({x \over y}) \otimes q_-(y)  \nonumber \\
+P_{++}^{qg} ({x \over y}) \otimes g_+(y)+
P_{+-}^{qg} ({x \over y}) \otimes g_-(y)),
\nonumber \\
{dq_-(x) \over{dt}}=
{\alpha_s \over {2 \pi}} (P_{+-} ({x \over y}) \otimes q_+(y)+
P_{++} ({x \over y}) \otimes q_-(y) \nonumber \\
+P_{+-}^{qg} ({x \over y}) \otimes g_+(y)+
P_{++}^{qg} ({x \over y}) \otimes g_-(y)),  \nonumber \\
{dg_+(x) \over{dt}}=
{\alpha_s \over {2 \pi}} (P_{++}^{gq} ({x \over y}) \otimes q_+(y)+
P_{+-}^{gq} ({x \over y}) \otimes q_-(y) \nonumber \\
+P_{++}^{gg} ({x \over y}) \otimes g_+(y)+
P_{+-}^{gg} ({x \over y}) \otimes g_-(y)),  \nonumber \\
{dg_-(x) \over{dt}}=
{\alpha_s \over {2 \pi}} (P_{+-}^{gq} ({x \over y}) \otimes q_+(y)+
P_{++}^{gq} ({x \over y}) \otimes q_-(y) \nonumber \\
+P_{+-}^{g} ({x \over y}) \otimes g_+(y)+
P_{++}^{gg} ({x \over y}) \otimes g_-(y)).
\end{eqnarray}
The non-singlet (valence) analogue of this equation is also easy to
write down
\begin{eqnarray}
{dq_{+, V}(x) \over{dt}}=
{\alpha_s \over {2 \pi}} (P_{++} ({x \over y}) \otimes q_{+,V}(y)+
P_{+-} ({x \over y}) \otimes q_{-,V}(y)), \nonumber \\
{dq_{-,V}(x) \over{dt}}=
{\alpha_s \over {2 \pi}} (P_{+-} ({x \over y}) \otimes q_{+,V}(y)+
P_{++} ({x \over y}) \otimes q_{-,V}(y)).
\end{eqnarray}
where the $q_{\pm,V}=q_\pm - \bar{q}_\pm$ are the valence components of fixed helicity.
The kernels in this basis are given by 
\beqa
P_{NS\pm,++}^{(0)} &=&P_{qq, ++}^{(0)}=P_{qq}^{(0)}\nonumber \\
P_{qq,+-}^{(0)}&=&P_{qq,-+}^{(0)}= 0\nonumber \\
P_{qg,++}^{(0)}&=& n_f x^2\nonumber \\
P_{qg,+-}&=& P_{qg,-+}= n_f(x-1)^2 \nonumber \\
P_{gq,++}&=& P_{gq,--}=C_F\frac{1}{x}\nonumber \\ 
P_{gg,++}^{(0)}&=&P_{gg,++}^{(0)}= N_c
\left(\frac{2}{(1-x)_+} +\frac{1}{x} -1 -x - x^2 \right) +{\beta_0}\delta(1-x) \nonumber \\
P_{gg,+-}^{(0)}&=& N_c
\left( 3 x +\frac{1}{x} -3 - x^2 \right). 
\eeqa

Taking linear combinations of these equations (adding and subtracting),
one recovers the usual evolutions for unpolarized $q(x)$ and longitudinally
polarized $\Delta q(x)$ distributions.

\section{The mathematical structure of the kernel}
Here we try to illustrate some simple manipulations on the kernels which are very useful 
in order to simplify the equations. Their typical form 
is the following\begin{equation}
P(x)=P_{1}(x)+\frac{P_{2}(x)}{(1-x)_{+}}+P_{3}\delta(1-x),\label{eq:general_kernel}\end{equation}
with a regular part $P_{1}(x)$, a {}``plus distribution'' part
$P_{2}(x)/(1-x)_{+}$ and a part $P_{3}$ which multiplies a Dirac delta function.
For a generic function $\alpha(x)$ defined in the $[0,1)$ interval
and singular in $x=1$, the plus distribution $[\alpha(x)]_{+}$ is
defined by\begin{equation}
\int_{0}^{1}f(x)[\alpha(x)]_{+}\textrm{d}x=\int_{0}^{1}\left(f(x)-f(1)\right)\alpha(x)\textrm{d}x,\end{equation}
where $f(x)$ is a regular test function. Alternatively, an operative
definition (in the sense of distributions)
is the following\begin{equation}
[\alpha(x)]_{+}=\alpha(x)-\delta(1-x)\int_{0}^{1}\alpha(y)\textrm{d}y.\label{eq:def_plus}\end{equation}
From (\ref{eq:def_plus}) it follows immediately that each plus distribution
integrate to zero in the $[0,1]$ interval\begin{equation}
\int_{0}^{1}[\alpha(x)]_{+}\textrm{d}x=0.\end{equation}

We want to make the convolution of the generic kernel (\ref{eq:general_kernel})
with a function $f(x)$.  We introduce the
notation $\bar{f}(x)=xf(x)$, the factor $x$ being introduced to stabilize the evolution for small 
$x$ values. 
The treatment of the regular and the
delta-function parts is trivial\begin{equation}
P_{1}(x)\otimes\bar{f}(x)=xP_{1}(x)\otimes f(x)=x\int_{x}^{1}\frac{\textrm{d}y}{y}P_{1}(y)f\left(\frac{x}{y}\right)=\int_{x}^{1}\textrm{d}yP_{1}(y)\bar{f}\left(\frac{x}{y}\right)\label{eq:conv_1}\end{equation}
\begin{equation}
P_{3}\delta(1-x)\otimes\bar{f}(x)=\int_{x}^{1}\frac{\textrm{d}y}{y}P_{3}\delta(1-y)\bar{f}\left(\frac{x}{y}\right).\label{eq:conv_3}\end{equation}

Let us now treat the more involved case of the plus distribution part\begin{eqnarray}
\frac{P_{2}(x)}{(1-x)_{+}}\otimes f(x) & = & \frac{P_{2}(x)}{1-x}\otimes f(x)-\left(\int_{0}^{1}\frac{\textrm{d}y}{1-y}\right)P_{2}(x)\delta(1-x)\otimes f(x)\nonumber \\
 & = & \int_{x}^{1}\frac{\textrm{d}y}{y}\frac{P_{2}(y)}{1-y}f\left(\frac{x}{y}\right)-\int_{0}^{1}\frac{\textrm{d}y}{1-y}\int_{x}^{1}\frac{\textrm{d}y}{y}P_{2}(y)\delta(1-y)f\left(\frac{x}{y}\right)\nonumber \\
 & = & \int_{x}^{1}\frac{\textrm{d}y}{y}\frac{P_{2}(y)}{1-y}f\left(\frac{x}{y}\right)-P_{2}(1)f(x)\int_{0}^{1}\frac{\textrm{d}y}{1-y}\nonumber \\
 & = & \int_{x}^{1}\frac{\textrm{d}y}{y}\frac{P_{2}(y)}{1-y}f\left(\frac{x}{y}\right)-P_{2}(1)f(x)\int_{x}^{1}\frac{\textrm{d}y}{1-y}\nonumber \\
 &  & \qquad\qquad\qquad\qquad\,\,\,-P_{2}(1)f(x)\int_{0}^{x}\frac{\textrm{d}y}{1-y}\\
 & = & \int_{x}^{1}\frac{\textrm{d}y}{y}\frac{P_{2}(y)f(x/y)-yP_{2}(1)f(x)}{1-y}+f(x)\log(1-x),\end{eqnarray}
which yields\begin{equation}
\frac{P_{2}(x)}{(1-x)_{+}}\otimes\bar{f}(x)=\int_{x}^{1}\textrm{d}y\frac{P_{2}(y)\bar{f}(x/y)-P_{2}(1)\bar{f}(x)}{1-y}+\bar{f}(x)\log(1-x).\label{eq:conv_2}\end{equation}

From Feynman diagrams calculations one can get just the regular part
$P_{1}(x)$ of each kernel. The remaining distributional parts (plus
distribution and delta distribution) emerge from a procedure of regularization,
that introduce the plus distribution part to regularize the eventual
singularity in $x=1$ and the delta distribution to fulfil some physical
constraints, the \emph{sum rules}.

The first one is the \emph{baryon number sum rule} (BNSR), asserting
that the baryon number (number of quarks less number of antiquarks)
of the hadron must remain equal to its initial value (3 in the case
of the proton) throughout the evolution, i.e.~for each value of $Q^{2}$\begin{equation}
q_{1}^{(-)}(Q^{2})=\int_{0}^{1}q^{(-)}(x,Q^{2})\textrm{d}x=3.\label{eq:BNSR}\end{equation}
Deriving (\ref{eq:BNSR}) with respect to $\log(Q^{2})$ and having
in mind that $q^{(-)}$ evolves with $P_{NS}^{V}$, we get\begin{equation}
\int_{0}^{1}\textrm{d}x\left[P_{NS}^{V}(Q^{2})\otimes q^{(-)}(Q^{2})\right](x)=0.\end{equation}
Making use of the property of the Mellin moment of a convolution (\ref{eq:Mellin_product})
this implies\begin{equation}
\left(\int_{0}^{1}P_{NS}^{V}(x,Q^{2})\textrm{d}x\right)\left(\int_{0}^{1}q^{(-)}(x,Q^{2})\textrm{d}x\right)=0,\end{equation}
from which, using (\ref{eq:BNSR}), we find the BNSR condition on
the kernel\begin{equation}
\int_{0}^{1}P_{NS}^{V}(x)\textrm{d}x=0.\label{eq:BNSR_kernel}\end{equation}

The other constraint is the \emph{momentum sum rule} (MSR), asserting
that the total momentum of the hadron is constant throughout the evolution.
Having in mind that $x$ is the fraction of momentum carried out by
each parton, this concept is translated into the relation\begin{equation}
\int_{0}^{1}\left(xq^{(+)}(x,Q^{2})+xg(x,Q^{2})\right)\textrm{d}x=1\label{eq:MSR}\end{equation}
that must hold for each value of $Q^{2}$. Deriving with respect to
$\log(Q^{2})$ and using the singlet DGLAP equation one obtains
\begin{eqnarray}
\int_{0}^{1}\textrm{d}x\, x\left\{ \left[P_{qq}(Q^{2})\otimes q^{(+)}(Q^{2})\right](x)+\left[P_{qg}(Q^{2})\otimes g(Q^{2})\right](x)\right.\nonumber \\
\left.+\left[P_{gq}(Q^{2})\otimes q^{(+)}(Q^{2})\right](x)+\left[P_{gg}(Q^{2})\otimes g(Q^{2})\right](x)\right\}  & = & 0.\end{eqnarray}
In a similar fashion, using (\ref{eq:Mellin_product}) we also obtain
\begin{eqnarray}
\left[\int_{0}^{1}x\left(P_{qq}(x,Q^{2})+P_{gq}(x,Q^{2})\right)\textrm{d}x\right]\left[\int_{0}^{1}xq^{(+)}(x,Q^{2})\textrm{d}x\right]\nonumber \\
+\left[\int_{0}^{1}x\left(P_{qg}(x,Q^{2})+P_{gg}(x,Q^{2})\right)\textrm{d}x\right]\left[\int_{0}^{1}xg(x,Q^{2})\textrm{d}x\right] & = & 0,\end{eqnarray}
from which we find the MSR conditions on the singlet kernels\begin{equation}
\int_{0}^{1}x\left(P_{qq}(x,Q^{2})+P_{gq}(x,Q^{2})\right)\textrm{d}x=0,\label{eq:MSR1_kernel}\end{equation}
\begin{equation}
\int_{0}^{1}x\left(P_{qg}(x,Q^{2})+P_{gg}(x,Q^{2})\right)\textrm{d}x=0.\label{eq:MSR2_kernel}\end{equation}

\subsection{The kernels and their regularization}

We illustrate now an example of the regularization procedure of the DGLAP kernels through the sum rules. The LO kernels computed
by diagrammatic techniques for $x<1$ are\begin{equation}
P_{qq}^{(0)}(x)=P_{NS}^{(0)}(x)=C_{F}\left[\frac{1+x^{2}}{1-x}\right]=C_{F}\left[\frac{2}{1-x}-1-x\right]\end{equation}
\begin{equation}
P_{qg}^{(0)}(x)=2T_{f}\left[x^{2}+(1-x)^{2}\right]\end{equation}
\begin{equation}
P_{gq}^{(0)}(x)=C_{F}\left[\frac{1+(1-x)^{2}}{x}\right]\end{equation}
\begin{equation}
P_{gg}^{(0)}(x)=2N_{C}\left[\frac{1}{1-x}+\frac{1}{x}-2+x(1-x)\right].\end{equation}
We want to analytically continue these kernels to $x=1$ curing the
ultraviolet singularities in $P_{qq}^{(0)}(x)$ and $P_{gg}^{(0)}(x)$.
We start introducing the plus distribution prescription in $P_{qq}^{(0)}(x)$.
We make the replacement\begin{equation}
\frac{1}{1-x}\longrightarrow\frac{1}{(1-x)_{+}}\end{equation}
to avoid the singularity and we add a term $k\delta(1-x)$ (where
$k$ has to be determined) to fulfill the BNSR (\ref{eq:BNSR_kernel}).
So we have\begin{equation}
P_{qq}^{(0)}(x)\longrightarrow C_{F}\left[\frac{2}{(1-x)_{+}}-1-x+k\delta(1-x)\right].\end{equation}
Imposing by the BNSR that $P_{qq}^{(0)}(x)$ integrates to zero in
$[0,1]$ and remembering that the plus distribution integrates to
zero we get\begin{equation}
\int_{0}^{1}P_{qq}^{(0)}(x)\textrm{d}x=C_{F}\left[-1-\frac{1}{2}+k\right]=0,\end{equation}
hence $k=3/2$, and the regularized form of the kernel is\begin{equation}
P_{qq}^{(0)}(x)=C_{F}\left[\frac{2}{(1-x)_{+}}-1-x+\frac{3}{2}\delta(1-x)\right].\end{equation}
Noticing that\begin{equation}
\int_{0}^{1}\frac{x}{(1-x)_{+}}\textrm{d}x=\int_{0}^{1}\frac{x-1+1}{(1-x)_{+}}\textrm{d}x=\int_{0}^{1}\left(-1+\frac{1}{(1-x)_{+}}\right)\textrm{d}x=-1\end{equation}
it can be easily proved that the MSR (\ref{eq:MSR1_kernel}) is satisfied.
Let us now regularize $P_{gg}(x)$. We make the replacement\begin{equation}
P_{gg}^{(0)}(x)\longrightarrow2N_{C}\left[\frac{1}{(1-x)_{+}}+\frac{1}{x}-2+x(1-x)\right]+k\delta(1-x).\end{equation}
 Imposing the other MSR (\ref{eq:MSR2_kernel}) we get\begin{eqnarray}
\int_{0}^{1}\left\{ 2N_{C}\left[\frac{x}{(1-x)_{+}}+1-2x+x^{2}(1-x)\right]+kx\delta(1-x)\right.\nonumber \\
\left.+2T_{f}\left[x^{3}+x(1-x)^{2}\right]\right\} \textrm{d}x & = & 0,\end{eqnarray}
from which we find\begin{equation}
k=\frac{11}{6}N_{C}-\frac{2}{3}T_{f}=\frac{\beta_{0}}{2},\end{equation}
so the regularized form of the kernel is\begin{equation}
P_{gg}^{(0)}(x)=2N_{C}\left[\frac{1}{(1-x)_{+}}+\frac{1}{x}-2+x(1-x)\right]+\frac{\beta_{0}}{2}\delta(1-x).\end{equation}

\section{First view: an ansatz from x-space and some examples}

In order to solve the evolution equations directly in \( x \)-space,
we assume solutions of the form\begin{equation}
\label{eq:ansatz}
f(x,Q^{2})=\sum _{n=0}^{\infty }\frac{A_{n}(x)}{n!}\log ^{n}\frac{\alpha _{s}(Q^{2})}{\alpha _{s}(Q_{0}^{2})}+\alpha _{s}(Q^{2})\sum _{n=0}^{\infty }\frac{B_{n}(x)}{n!}\log ^{n}\frac{\alpha _{s}(Q^{2})}{\alpha _{s}(Q_{0}^{2})},
\label{rossis}
\end{equation}
for each parton distribution \( f \), where $Q_0$ defines the initial 
evolution scale. The justification of this ansatz can be found, 
at least in the case of the photon structure function, 
in the original work of Rossi \cite{Rossi}, and its connection 
to the ordinary solutions of the DGLAP equations is most easily 
worked out by taking moments of the scale invariant coefficient 
functions $A_n$ and $B_n$ and comparing them to 
the corresponding moments 
of the parton distributions, as we are going to illustrate 
in section 5. The link between Rossi's expansion 
and the solution of the evolution equations 
(which are ordinary differential equations) in the space 
of the moments up to NLO will be discussed in that section, from which 
it will be clear that 
Rossi's ansatz involves a resummation 
of the ordinary Mellin moments of the parton distributions.   

Setting \( Q=Q_{0} \) in (\ref{eq:ansatz})
we get\begin{equation}
\label{eq:boundary}
f(x,Q_{0}^{2})=A_{0}(x)+\alpha _{s}(Q_{0}^{2})B_{0}(x).
\end{equation}
Inserting (\ref{eq:ansatz}) in the evolution equations, we obtain
the following recursion relations for the coefficients \( A_{n} \)
and \( B_{n} \) 
\begin{equation}
\label{eq:An_recurrence}
A_{n+1}(x)=-\frac{2}{\beta _{0}}P^{(0)}(x)\otimes A_{n}(x),
\end{equation}

\begin{equation}
B_{n+1}(x)=-B_{n}(x)-\frac{\beta_{1}}{4\pi\beta_{0}}A_{n+1}(x)-\frac{2}{\beta_{0}}P^{(0)}(x)\otimes B_{n}(x)-\frac{1}{\pi\beta_{0}}P^{(1)}(x)\otimes A_{n}(x) 
\label{eq:Bn_recurrence}
\end{equation}
obtained by equating left-hand sides and right-hand-side of the equation 
of the same logarithmic power 
in $\log^n\alpha_s(Q^2)$ and $\alpha_s \log^n \alpha_s(Q^2)$. 
Any boundary condition satisfying (\ref{eq:boundary}) can be chosen at the lowest scale $Q_0$ and in our case we choose\begin{equation}
\label{eq:initial}
B_{0}(x)=0,\qquad f(x,Q_{0}^{2})=A_{0}(x).
\end{equation}
In the numerical implementations of this procedure 
one has to be careful in the manipulations of the numerical integrals, being all the kernels defined as 
distributions. Since the distributions are integrated, 
there are various ways to render the integrals finite, 
as discussed in the previous literature on the method \cite{Storrow} 
in the case of the photon structure function. 
In these previous studies the 
edge-point contributions - i.e. the terms which multiply $\delta(1-x)$ 
in the kernels - are approximated using a sequence of functions 
converging to the $\delta$ function in a distributional sense.

This technique is not very efficient. We think that 
the best way to proceed is to actually perform the integrals explicitly in the recursion relations and let the subtracting 
terms appear under the same integral together with the 
bulk contributions ($x<1$). This procedure is best exemplified 
by the integral relation 
\beq
\int_x^1 \frac{dy}{y (1-y)_+}f(x/y)=\int_x^1\frac{dy}{y}
\frac{ yf(y) - x f(x)}{y-x} -\log(1-x) f(x)
\label{simplerel}
\eeq
in which, on the right hand side, regularity of both the first 
and the second term is explicit. For instance, the evolution equations become 
(prior to separation between singlet and non-singlet sectors) in the unpolarized 
case
\beqa
\frac{d q_i(x)}{d \log(Q^2)} &=& 2 C_F 
\int\frac{dy}{y}\frac{ y q_i(y) - x q_i(x)}{y-x} +2 C_F \log(1-x) q_i(x) -
\int_x^1\frac{dy}{y}\left( 1 + z\right)q_i(y) + 
\frac{3}{2} C_F q(x) \nonumber \\
&& + n_f\int_x^1\frac{dy}{y}
\left( z^2 +(1-z)^2\right)g(y)\nonumber \\
\frac{d g(x)}{d \log(Q^2)} &=& 
C_F \int_x^1\frac{dy}{y}\frac{1 +(1-z)^2}{z}q_i(y)
+ 2 N_c \int_x^1\frac{dy}{y}
\frac{ y f(y) - x f(x)}{y-x}g(y) 
\nonumber \\
&& + 2 N_c \log(1-x) g(x) 
+2 N_c\int_x^1 \frac{dy}{y}\left( \frac{1}{z} -2 + z(1-z)\right)g(y) + 
\frac{\beta_0}{2}g(x) \nonumber \\
\eeqa
with $z\equiv x/y$. Here $q$ are fixed flavour distributions.

\section{The evolution of the transverse spin distributions}
We show here an example, worked out in some detail, that illustrates the kind 
of manipulations that are involved in the computation of the recursion 
relations. 

In fact, leading order (LO) and NLO recursion relations for the coefficients of the expansion 
can be worked out quite easily. We illustrate here in detail 
the implementation of a non-singlet evolutions, such 
as those involving transverse spin distributions. 
For the first recursion relation (\ref{eq:An_recurrence}) in this case 
we have
\beqn
&&A^{\pm}_{n+1}(x)=-\frac{2}{\beta_{0}}\Delta_{T}P^{(0)}_{qq}(x)\otimes A^{\pm}_{n}(x)=\nonumber\\ 
&&C_{F}\left(-\frac{4}{\beta_{0}}\right)\left[\int^{1}_{x}\frac{dy}{y}\frac{y A^{\pm}_{n}(y) - x A^{\pm}_{n}(x)}{y-x} + A^{\pm}_{n}(x) \log(1-x)\right]+\nonumber\\
&&C_{F}\left(\frac{4}{\beta_{0}}\right) \left(\int_{x}^{1}\frac{dy}{y} A^{\pm}_{n}(y)\right) + C_{F}\left(-\frac{2}{\beta_{0}}\right)\frac{3}{2} A^{\pm}_{n}(x)\,.
\eeqn
As we move to NLO, it is convenient to summarize 
the structure of the transverse kernel $\Delta_{T}P^{\pm, (1)}_{qq}(x)$ 
\cite{Vogelsang} as  

\beqn
&&\Delta_{T}P^{\pm, (1)}_{qq}(x)= K^{\pm}_{1}(x)\delta(1-x) + K^{\pm}_{2}(x)S_{2}(x) +K^{\pm}_{3}(x)\log(x)\nonumber\\
&&+ K^{\pm}_{4}(x)\log^{2}(x) +K^{\pm}_{5}(x)\log(x)\log(1-x) + K^{\pm}_{6}(x)\frac{1}{(1-x)_{+}} + K^{\pm}_{7}(x)\,.    
\eeqn

Hence, for the $(+)$ case we have 

\beqn
&&\Delta_{T}P^{+, (1)}_{qq}(x)\otimes A^{+}_{n}(x) = K^{+}_{1} A^{+}_{n}(x) + \int^{1}_{x}\frac{dy}{y}\left[K^{+}_{2}(z) S_{2}(z) + K^{+}_{3}(z)\log(z) \right.\nonumber\\
&& \left. + \log^{2}(z)K^{+}_{4}(z) + \log(z)\log(1-z)K^{+}_{5}(z)\right] A^{+}_{n}(y) +  \nonumber\\ 
&&K^{+}_{6}\left\{\int^{1}_{x}\frac{dy}{y} \frac{yA^{+}_{n}(y) - xA^{+}_{n}(x)}{y-x} + A^{+}_{n}(x)\log(1-x) \right\} + K^{+}_{7}\int^{1}_{x}\frac{dy}{y}A^{+}_{n}(y)\,, 
\eeqn

where $z={x}/{y}$. For the $(-)$ case we get a similar expression.
  
For the $B^{\pm}_{n+1}(x)$  we get (for the $(+)$ case) 

\ba
&&B^{+}_{n+1}(x) = - B^{+}_{n}(x) + \frac{\beta_{1}}{2\beta^{2}_{0}} \left\{2C_{F}\left[\int^{1}_{x}\frac{dy}{y}\frac{y A^{+}_{n}(y) - x A^{+}_{n}(x)}{y-x} + A^{+}_{n}(x) \log(1-x)\right]\right.+\nonumber\\
&&\left.-2C_{F}\left(\int_{x}^{1}\frac{dy}{y} A^{+}_{n}(y)\right) + C_{F}\frac{3}{2} A^{+}_{n}(x)\right\}-\frac{1}{4\pi\beta_{0}}K^{+}_{1} A^{+}_{n}(x)+ \int^{1}_{x}\frac{dy}{y}\left[ K^{+}_{2}(z) S_{2}(z) + \right.\nonumber\\
&&+ \left.K^{+}_{3}(z)\log(z)+\log^{2}(z)K^{+}_{4}(z) + \log(z)\log(1-z)K^{+}_{5}(z)\right]\left(-\frac{1}{4\pi\beta_{0}}\right)A^{+}_{n}(y)+\nonumber\\
&&K^{+}_{6}\left(-\frac{1}{4\pi\beta_{0}}\right)\left\{\left[\int^{1}_{x}\frac{dy}{y} \frac{yA^{+}_{n}(y) - xA^{+}_{n}(x)}{y-x} + A^{+}_{n}(x)\log(1-x) \right] + K^{+}_{7}\int^{1}_{x}\frac{dy}{y}A^{+}_{n}(y)\right\}-\nonumber\\
&&C_{F}\left(-\frac{4}{\beta_{0}}\right)\left[\int^{1}_{x}\frac{dy}{y}\frac{y B^{\pm}_{n}(y) - x B^{\pm}_{n}(x)}{y-x} + B^{\pm}_{n}(x) \log(1-x)\right]+\nonumber\\
&&C_{F}\left(\frac{4}{\beta_{0}}\right) \left(\int_{x}^{1}\frac{dy}{y} B^{\pm}_{n}(y)\right) + C_{F}\left(-\frac{2}{\beta_{0}}\right)\frac{3}{2} B^{\pm}_{n}(x).\nonumber
\ea
The explicit expressions of the $K_i$ can be found in \cite{CPC}.

\subsection{Solutions in moments space and comparisons}
It is interesting to compare the recursive solution with the solution in moment space, that is easy to derive. In moment space the equations become 
algebraic and can be readily solved. For this we need some definitions. 
   
The $n$-th Mellin moment of a function of the Bjorken variable $f(x)$
is defined by\begin{equation}
f_{n}=\int_{0}^{1}x^{n-1}f(x)\textrm{d}x.\end{equation}
An important property of Mellin moments is that the Mellin moment
of the convolution of two functions is equal to the product of the
individual Mellin moments\begin{equation}
\left[f\otimes g\right]_{n}=f_{n}g_{n}.\label{eq:Mellin_product}\end{equation}
Let us prove it.\begin{equation}
\left[f\otimes g\right]_{n}=\int_{0}^{1}\textrm{d}x\, x^{n-1}\left[f\otimes g\right](x)=\int_{0}^{1}\textrm{d}x\, x^{n-1}\int_{x}^{1}\frac{\textrm{d}y}{y}f(y)g\left(\frac{x}{y}\right).\end{equation}
Exchanging the $x$ and $y$ integrations \begin{equation}
\left[f\otimes g\right]_{n}=\int_{0}^{1}\textrm{d}y\, f(y)\int_{0}^{y}\frac{\textrm{d}x}{y}x^{n-1}g\left(\frac{x}{y}\right),\end{equation}
and introducing the new variable $z=x/y$\begin{equation}
\left[f\otimes g\right]_{n}=\int_{0}^{1}\textrm{d}y\, f(y)y^{n-1}\int_{0}^{1}\textrm{d}z\, 
z^{n-1}g(z)=f_{n}g_{n}.\end{equation}

This leads to an alternative formulation of DGLAP equation, that is
also the most widely used to solve numerically the evolution equations.
By taking the first Mellin moment of both sides of the integro-differential
equation (\ref{eq:DGLAP}) we are left with the differential equation\begin{equation}
\frac{\textrm{d}}{\textrm{d}\log Q^{2}}f_{1}(Q^{2})=P_{1}(Q^{2})f_{1}(Q^{2})\end{equation}
that can be easily solved to give\begin{equation}
f_{1}(Q^{2})=\int_{0}^{1}f(x,Q^{2})\textrm{d}x.\end{equation}
To get the desired solution $f(x,Q^{2})$ there is a last step, the
inverse Mellin transform of the first moment of the parton distributions,
involving a numerical integration on the complex plane. This is the
most difficult (and time-consuming) task that the algorithms of solution
of DGLAP equation based on Mellin transformation -- by far the most
widely used -- must accomplish. 
It is particularly instructing to illustrate here briefly the relation 
between the Mellin moments of the parton distributions, which evolve 
with standard ordinary differential equations, and those of the 
arbitrary coefficient $A_n(x)$ and $B_n(x)$ which characterize 
Rossi's expansion up to next-to leading order (NLO). This relation, as we are going to show, involves a 
resummation of the ordinary moments of the parton distributions. 

Specifically, here we will be dealing with the relation between 
the Mellin moments of the coefficients appearing in the expansion 
\beqa
A(N) &=& \int_0^1\,dx \, x^{N-1} A(x)\nonumber \\
B(N) &=&\int_0^1\,dx \, x^{N-1} B(x) \nonumber \\
\eeqa
and those of the distributions  
\beq
\Delta_T q^{(\pm)}(N,Q^2)=\int_0^1\,dx \, x^{N-1} \Delta_T q^{(\pm)}(x,Q^2)).
\eeq 
For this purpose we recall that the general (non-singlet) solution to NLO for the latter moments is given by 
\begin{eqnarray} \label{evsol}
\nonumber
\Delta_T q_{\pm} (N,Q^2) &=& K(Q_0^2,Q^2,N)
\left( \frac{\alpha_s (Q^2)}{\alpha_s (Q_0^2)}\right)^{-2\Delta_T 
P_{qq}^{(0)}(N)/ \beta_0}\! \Delta_T q_{\pm}(N, Q_0^2)
\label{solution}
\end{eqnarray}
with the input distributions $\Delta_T q_{\pm}^n (Q_0^2)$ at the input scale 
$Q_0$.
We also have set 
\beq
K(Q_0^2,Q^2,N)= 1+\frac{\alpha_s (Q_0^2)-
\alpha_s (Q^2)}{\pi\beta_0}\!
\left[ \Delta_T P_{qq,\pm}^{(1)}(N)-\frac{\beta_1}{2\beta_0} \Delta_T 
P_{qq\pm}^{(0)}(N) \right]. 
\eeq
In the expressions above we have introduced the corresponding moments for the LO and NLO kernels 
($\Delta_T P_{qq}^{(0),N}$,
$ \Delta_T P_{qq,\pm}^{(1),N})$. 

The relation between the moments of the coefficients of the non-singlet
$x-$space expansion and those of the parton distributions at any $Q$, as expressed by eq.~(\ref{solution}) can be easily written down
\beq
A_n(N) + \alpha_s B_n(N)=\Delta_T q_\pm(N,Q_0^2)K(Q_0,Q,N)\left(\frac{-2 \Delta_T P_{qq}(N)}{\beta_0}\right)^n.
\label{relation}
\eeq

As a check of this expression, notice that the initial condition is easily obtained from  
(\ref{relation}) setting $Q\to Q_0, n\to 0$, thereby obtaining 
\beq
A_0^{NS}(N) + \alpha_s B_0^{NS} (N)= \Delta_T q_\pm(N,Q_0^2),
\eeq
which can be solved with $A_0^{NS}(N)=\Delta_T q_\pm(N,Q_0^2)$ and 
$B_0^{NS} (N)=0$. 

It is then evident that the expansion (\ref{rossis}) involves a resummation of the logarithmic contributions, as shown in eq.~(\ref{relation}). 

In the singlet sector we can work out a similar relation both to LO

\beq
A_n(N) = e_1\left(\frac{-2 \lambda_1}{\beta_0}\right)^n 
+e_2 \left(\frac{-2 \lambda_2}{\beta_0}\right)^n 
\eeq

with 
\beqa
e_1 &=& \frac{1}{\lambda_1 - \lambda_2}\left( P^{(0)}(N)- \lambda_2 \one \right)
\nonumber \\
e_2 &=& \frac{1}{\lambda_2 - \lambda_1}\left( - P^{(0)}(N) + \lambda_1 \one\right)
\nonumber \\
\lambda_{1,2}&=& \frac{1}{2}\left( 
P^{(0)}_{qq}(N) + P^{(0)}_{gg}(N) \pm \sqrt{\left(P^{(0)}_{qq}(N)- P^{(0)}_{gg}(N)\right)^2  
+ 4 P^{(0)}_{qg}(N)P^{(0)}_{gq}(N)}\right),
\eeqa
and to NLO 
\beq
A_n(N) + \alpha_s B_n(N) = \chi_1\left(\frac{-2 \lambda_1}{\beta_0}\right)^n 
+\chi_2 \left(\frac{-2 \lambda_2}{\beta_0}\right)^n, 
\eeq

where
\beqa
\chi_1 &=& e_1 + \frac{\alpha}{2 \pi}\left( \frac{-2}{\beta_0}e_1 \R e_1
+\frac{ e_2 \R e_1}{\lambda_1 - \lambda_2 - \beta_0/2}\right) \nonumber \\
\chi_2 &=& e_2 + \frac{\alpha}{2 \pi}\left( \frac{-2}{\beta_0}e_2 \R e_2
+\frac{ e_1 \R e_2}{\lambda_2 - \lambda_1 - \beta_0/2}\right)\nonumber \\
\eeqa
with
\beq
\R= P^{(1)}(N) -\frac{\beta_1}{2 \beta_0}P^{(0)}(N).
\eeq
We remark, if not obvious, that $A_n(N)$ and $B_n(N)$, $P^{(0)}(N)$, $P^{(1)}(N)$, in this case,
are all 2-by-2 singlet matrices.

% R(x) &=& P^{(1)}(x)-{\beta_1\over 2 \beta_0} P^{(0)}(x).

\section{Moving to higher orders} 
The structure of the solution to higher orders can be worked out in generality. More 
details can be found in \cite{Cacorguz}, here we just outline the procedure. 

The perturbative expansion of the kernels now includes the NNLO contributions 
and is given by \begin{equation}
P(x,\alpha_{s})=\left(\frac{\alpha_{s}}{2\pi}\right)P^{(0)}(x)+\left(\frac{\alpha_{s}}{2\pi}\right)^{2}P^{(1)}(x)+\left(\frac{\alpha_{s}}{2\pi}\right)^{3}P^{(2)}(x)+\ldots.\label{eq:kernel_expansion}\end{equation} 
whose specific form can be found in the original literature \cite{NNLO_nonsinglet,NNLO_singlet}. 

We solve Eq. (\ref{eq:DGLAP}) directly in $x$-space, 
assuming a solution of the form\begin{eqnarray}
f(x,Q^{2}) & = & \sum_{n=0}^{\infty}\frac{A_{n}(x)}{n!}\log^{n}\frac{\alpha_{s}(Q^{2})}{\alpha_{s}(Q_{0}^{2})}+\alpha_{s}(Q^{2})\sum_{n=0}^{\infty}\frac{B_{n}(x)}{n!}\log^{n}\frac{\alpha_{s}(Q^{2})}{\alpha_{s}(Q_{0}^{2})}\nonumber \\
 &  & +\left(\alpha_{s}(Q^{2})\right)^{2}\sum_{n=0}^{\infty}\frac{C_{n}(x)}{n!}\log^{n}\frac{\alpha_{s}(Q^{2})}{\alpha_{s}(Q_{0}^{2})}\label{eq:ansatz2}\end{eqnarray}
for each parton distribution $f$, where $Q_{0}$ defines the initial
evolution scale. 

As in the previous examples, also in this case we derive the following recursion relations for the coefficients
$A_{n}$, $B_{n}$ and $C_{n}$ 
\begin{equation}
A_{n+1}(x)=-\frac{2}{\beta_{0}}P^{(0)}(x)\otimes A_{n}(x),\label{eq:An_recurrence2}\end{equation}
\begin{equation}
B_{n+1}(x)=-B_{n}(x)-\frac{\beta_{1}}{4\pi\beta_{0}}A_{n+1}(x)-\frac{2}{\beta_{0}}P^{(0)}(x)\otimes B_{n}(x)-\frac{1}{\pi\beta_{0}}P^{(1)}(x)\otimes A_{n}(x),\label{eq:Bn_recurrence2}\end{equation}

\begin{eqnarray}
C_{n+1}(x) & = & -2C_{n}(x)-\frac{\beta_{1}}{4\pi\beta_{0}}B_{n}(x)-\frac{\beta_{1}}{4\pi\beta_{0}}B_{n+1}(x)-\frac{\beta_{2}}{16\pi^{2}\beta_{0}}A_{n+1}(x)\nonumber \\
 &  & -\frac{2}{\beta_{0}}P^{(0)}(x)\otimes C_{n}(x)-\frac{1}{\pi\beta_{0}}P^{(1)}(x)\otimes B_{n}(x)\nonumber \\
 &  & -\frac{1}{2\pi^{2}\beta_{0}}P^{(2)}(x)\otimes A_{n}(x).\label{eq:Cn_recurrence}\end{eqnarray}
It is an easy exercise to derive the recursion relations for the coefficients of the expansion. We illustrate the derivation for the interested reader. 

We introduce the notation\begin{equation}
L(Q^{2})=\log\frac{\alpha_{s}(Q^{2})}{\alpha_{s}(Q_{0}^{2})}\end{equation}
and, making use of the beta function definition (\ref{eq:beta_def}),
we compute its derivative\begin{equation}
\frac{\textrm{d}L(Q^{2})}{\textrm{d}\log Q^{2}}=\frac{\alpha_{s}(Q_{0}^{2})}{\alpha_{s}(Q^{2})}\frac{\textrm{d}}{\textrm{d}\log Q^{2}}\frac{\alpha_{s}(Q^{2})}{\alpha_{s}(Q_{0}^{2})}=\frac{1}{\alpha_{s}(Q^{2})}\frac{\textrm{d}\alpha_{s}(Q^{2})}{\textrm{d}\log Q^{2}}=\frac{\beta(\alpha_{s})}{\alpha_{s}(Q^{2})}\end{equation}
Inserting our ansatz (\ref{eq:ansatz2}) for the solution into the
DGLAP equation (\ref{eq:DGLAP}) we get for the LHS\begin{eqnarray}
 &  & \sum_{n=1}^{\infty}\left\{ \frac{A_{n}(x)}{n!}nL^{n-1}\frac{\beta(\alpha_{s})}{\alpha_{s}}+\alpha_{s}\frac{B_{n}(x)}{n!}nL^{n-1}\frac{\beta(\alpha_{s})}{\alpha_{s}}\right.\nonumber \\
 &  & \left.\qquad+\alpha_{s}^{2}\frac{C_{n}(x)}{n!}nL^{n-1}\frac{\beta(\alpha_{s})}{\alpha_{s}}\right\} \nonumber \\
 &  & +\sum_{n=0}^{\infty}\left\{ \beta(\alpha_{s})\frac{B_{n}(x)}{n!}L^{n}+2\alpha_{s}\beta(\alpha_{s})\frac{C_{n}(x)}{n!}L^{n}\right\} .\end{eqnarray}
Sending $n\rightarrow n-1$ in the first sum, using the three-loop expansion
of the beta function (\ref{eq:beta_exp}) and neglecting all the terms
of order $\alpha_{s}^{4}$ or more, the previous formula becomes\begin{eqnarray}
 &  & \sum_{n=0}^{\infty}\left\{ \frac{A_{n+1}(x)}{n!}L^{n}\left(-\frac{\beta_{0}}{4\pi}\alpha_{s}-\frac{\beta_{1}}{16\pi^{2}}\alpha_{s}^{2}-\frac{\beta_{2}}{64\pi^{3}}\alpha_{s}^{3}\right)\right.\nonumber \\
 &  & +\frac{B_{n+1}(x)}{n!}L^{n}\left(-\frac{\beta_{0}}{4\pi}\alpha_{s}^{2}-\frac{\beta_{1}}{16\pi^{2}}\alpha_{s}^{3}\right)+\frac{C_{n+1}(x)}{n!}L^{n}\left(-\frac{\beta_{0}}{4\pi}\alpha_{s}^{3}\right)\nonumber \\
 &  & \left.+\frac{B_{n}(x)}{n!}L^{n}\left(-\frac{\beta_{0}}{4\pi}\alpha_{s}^{2}-\frac{\beta_{1}}{16\pi^{2}}\alpha_{s}^{3}\right)+2\frac{C_{n}(x)}{n!}L^{n}\left(-\frac{\beta_{0}}{4\pi}\alpha_{s}^{3}\right)\right\} .\label{eq:recrelLHS}\end{eqnarray}
Using the expansion of the kernels (\ref{eq:kernel_expansion}), we get for
the RHS\begin{eqnarray}
 &  & \sum_{n=0}^{\infty}\frac{L^{n}}{n!}\left\{ \frac{\alpha_{s}}{2\pi}\left[P^{(0)}\otimes A_{n}\right](x)+\frac{\alpha_{s}^{2}}{4\pi^{2}}\left[P^{(1)}\otimes A_{n}\right](x)\right.\nonumber \\
 &  & \quad\quad\quad+\frac{\alpha_{s}^{3}}{8\pi^{3}}\left[P^{(2)}\otimes A_{n}\right](x)+\frac{\alpha_{s}^{2}}{2\pi}\left[P^{(0)}\otimes B_{n}\right](x)\nonumber \\
 &  & \left.\quad\quad\quad+\frac{\alpha_{s}^{3}}{4\pi^{2}}\left[P^{(1)}\otimes B_{n}\right](x)+\frac{\alpha_{s}^{3}}{2\pi}\left[P^{(0)}\otimes C_{n}\right](x)\right\} .\label{eq:recrelRHS}\end{eqnarray}
Equating (\ref{eq:recrelLHS}) and (\ref{eq:recrelRHS}) term by term
and grouping the terms proportional respectively to $\alpha_{s}$,
$\alpha_{s}^{2}$ and $\alpha_{s}^{3}$ we get the three desired recursion
relations (\ref{eq:An_recurrence2}), (\ref{eq:Bn_recurrence2}) and
(\ref{eq:Cn_recurrence}).

Setting $Q=Q_{0}$ in (\ref{eq:ansatz2}) we get\begin{equation}
f(x,Q_{0}^{2})=A_{0}(x)+\alpha_{s}(Q_{0}^{2})B_{0}(x)+\left(\alpha_{s}(Q^{2})\right)^{2}C_{0}(x).\label{eq:boundary2}\end{equation}
We solve these relations using a boundary condition 
satisfying (\ref{eq:boundary2}). In our case we choose\begin{equation}
B_{0}(x)=C_{0}(x)=0,\qquad f(x,Q_{0}^{2})=A_{0}(x).\end{equation}
The procedure studied in the previous section can be
generalized and applied to obtain solutions that retain higher order
logarithmic contributions in the NLO/NNLO singlet cases.
The same procedure is also the one that
has been implemented in all the existing codes for the singlet:
one has to truncate the equation and then try to reach the exact solution by a sufficiently
high number of iterates.  
These arguments can be extended to even higher orders.
We have recently shown \cite{Cacorguz} that the exact solution to all orders of the RGE's of the p.d.f.'s can be written down 
in close form by the introduction of a generalized logarithmic expansion that includes contributions of all powers of the 
strong coupling constant $\alpha_s$ and are summarized by the expression 

\beq
f(x,Q^2)=\sum_{n=0}^{\infty} \frac{1}{n!}C_n(x,\alpha_s)\log^n\left(\frac{\alpha_s(Q^2)}{\alpha_s(Q_0^2)}\right) 
\eeq
where the function 
\beq
C_n(x,\alpha_s,\alpha_0)=\sum_{k=0}^{\infty} \alpha_s^k A_n^{(k)}(x)
\eeq
is determined perturbatively by the recursione relations extracted from the RGE. The analysis is quite 
involved and can be found in  the original literature.   

\section{Other solutions in x-space: the Laguerre expansion}
Another possible way to solve the evolution equations is to use an operatorial formalism and observe that the distributions can be expanded in a suitable orthonormal 
basis (Laguerre polynomials). This method has had limited applications due to the difficulty to handle numerically the small-x behaviour of the algorithm 
that constructs the solution. The implementation of the method is discussed in \cite{CS}. 
One starts by writing the expression of the solution in operatorial form in terms of two singlet evolution operators
$E_{\pm}(t,x)$ and initial conditions
$\tilde{q}_{\pm}(x,t=0)\equiv \tilde{q}_{\pm}(x)$ as

\beq
{d\over dt} E_{\pm}=P_{\pm}\otimes E_{\pm},
\eeq
whose solutions are given by
\beqa
&& q_i^{(-)}(t,x)= E_{(-)}\otimes \tilde{q}_i^{(-)} \nonumber \\
&& \chi_i(t,x)=E_{(+)}\otimes \tilde\chi_i(x).
\eeqa
The singlet evolution for the matrix operator $E(t,x)$

\beqa
&& \left( \begin{array}{cc}
E_{FF} & E_{FG} \\
E_{GF} & E_{GG}
\end{array} \right)
\eeqa

\beq
{d E\over d t}= P\otimes E
\eeq

is solved similarly as

\beqa
&&\left( \begin{array}{c}
 q^{(+)}
(t,x)\\
G(t,x)
\end{array}\right)= E(t,x)\otimes
\left( \begin{array}{c}
\tilde{q}^{(+)}(x)\\
\tilde{G}(x)
\end{array}\right). \\
\eeqa

This method, proposed in \cite{FurmanskiPetronzio}, requires
an expansion of the splitting functions and of the parton distributions in the
basis of the Laguerre polynomials

\beq
L_n(y)=\sum_{k=0}^n \left(\begin{array}{c} n \\ k \end{array} \right)
(-1)^k {y^k\over k!}
\eeq

which satisfies the property of closure under a convolution

\beq
L_n(y)\otimes L_m(y)=L_{n+m}(y) - L_{n +m +1}(y).
\eeq
In order to improve the small-x behaviour of the algorithm, from now on,
the evolution is applied to the modified kernel $x P(x)$, which,
for simplicity, is still denoted as in all the equations above,
i.e. by $P(x)$. At a second step, the $ 0<x<1$ interval
is mapped into an infinite interval $0<y<\infty$ by a change of variable
$x=e^{-y}$ and all the integrations are performed in this last interval.
We start from the non-singlet case by defining
the Laguerre expansion of the kernels and the corresponding (Laguerre) moments to lowest order

\beqa
 P_V^{(0)}(y)&=&\sum_{n=0}^\infty P_n^{(0)} L_n(y), \nonumber \\
P_n^{(0)}&=&\int_0^\infty dy e^{-y} L_n(y)\,\, P^{(0)}(y)
\eeqa

and to NLO

\beq
R(y)=\sum_{n=0}^\infty R_n L_n(y).
\eeq
One defines also the difference of  moments

\beqa
 p_i^{(0)}&=&P_i^{(0)} -P_{i-1}^{(0)}\,\,\,\,\,\,\,(P_{-1}^{(0)}=0) \nonumber \\
r_i&=& R_i- R_{i-1} \,\,\,\,\,\, R_{-1}=0.
\eeqa

A similar expansion is set up for the evolution operators  $E(t,y)$

\beqa
E^{(0)}(t,y)&=&\sum_{n=0}^\infty E_n^{(0)}(t) L_n(y)\nonumber \\
E(t,y)&=&\sum_{n=0}^\infty E_n(t) L_n(y),
\eeqa

where all the information on the $t$ evolution is contained in the moments $E_n(t)$. 
The solution to NLO is expressed as

\beqa
 E_n (t)=E_n^{(0)}(t) -{2\over \beta_0}{\alpha(t)-\alpha(0)\over 2\pi}E_n^{(1)}(t),
\label{solve}
\eeqa

where
\beqa
&& E_n^{(0)}(t)=e^{P_0^{(0} t} \sum_{k=0}^n {A_n^{(k)} t^k\over k!} \\
&& E_n^{(1)}(t)=\sum_i^n r_{n-i} E_i^{(0)}(t), \\
\eeqa
and the coefficients $A_n^{(k)}$ are determined recursively from the moments of the
lowest order kernel $P^{(0)}$

\beqa
 A_n^{(0)}&=& 1 \nonumber \\
 A_n^{(k+1)}&=&\sum_{i=k}^{n-1}p_{n-i}^{(0)} A_i^{(k)} \hspace{1cm} (k=0,1,2,...,n-1).
\eeqa

In the singlet case one proceeds in a similar way. The solution is expressed in terms of a
2-by-2 matrix operator

\beq
E^{(0)}(t,y)=\sum_{n=0}^\infty E_n^{(0)}(t) L_n(y).
\eeq
The solution (at leading order) is written down in terms of 2 projection matrices and one
eigenvalue ($\lambda$) of the $P^{(0)}$ (matrix) kernel

\beqa
e_1={1\over \lambda}P^{(0)}, \hspace{2cm}
e_2={1\over \lambda}\left( -P^{(0)} +\lambda \bf{1}\right), \\
\eeqa
where
\beqa
\lambda&=&-({4\over 3} C_F +{2\over 3} n_f T_R),
\eeqa

in the form
\beqa
E^{(0)}_n(t)&=&\sum_{k=0}^n {t^k\over k!}\left( A_n^{(k)}+ B_n^{(k)}e^{\lambda t}\right).
\eeqa

The recursion relations which allow to build $A_n^{(k)}$ and $B_n^{(k)}$ are solved in two
steps as follows. One solves first for two sets of matrices $a_n^{(k)}$ and $b_n^{(k)}$  by the relations

\beqa
a_n^{(0)}&=&0 \nonumber \\
 a_n^{(k+1)}&=&\lambda e_1 a_n^{(k)}+\sum_{i=k}^{n-1}p_{n-i}^{(0)} a_i^{(k)}\nonumber \\
 b_n^{(0)}&=& 0 \nonumber \\
 b_n^{(k+1)}&=& -\lambda e_2 b_n^{(k)} +\sum_{i=1}^{n-1} p_{n-i}^{(0)} b_i^{(k)},
\label{solve1}
\eeqa

which are used to construct the matrices $A_n^{(0)}$ and $B_n^{(0)}$

\beqa
A_n^{(0)}&=& e_2 -{1\over \lambda^n}\left( e_1 a_n^{(n)} - (-1)^n e_2 b_n^{(n)}\right)\nonumber \\
B_n^{(0)}&=& e_1 +{1\over \lambda^n}\left( e_1 a_n^{(n)} - (-1)^n e_2 b_n^{(n)}\right) . \\
\eeqa

These matrices are then input in the recursion relations

\beqa
A_0^{(0)}&=& e_2 \hspace{1cm}B_0^{(0)}= e_1 \nonumber \\
A_n^{(k+1)}&=& \lambda e_1 A_n^{(k)} +\sum_{i=k}^{n-1}p_{n-i}^{(0)} A_i^k \nonumber \\
B_n^{k+1}&=& -\lambda e_2 B_n^{(k)}+\sum_{i=k}^{n-1} p_{n-i}^{(0)} B_i^{(k)}
\eeqa
witn $n>0$ and $k=0,1,...,n-1$,
which generates the coefficients of the matrix-valued operator $E^{(0)}$
(i.e. the leading order solution). The NLO part of the evolution is obtained from

\beqa
E^{(1)}(t,y)&=&\sum_{n=0}^{\infty} E_n^{(1)}(t) L_n(y),
\eeqa

with
\beqa
E_n^{(1)}(t)&=&\tilde{E}_{n}^{(1)}(t)-2 \tilde{E}_{n-1}^{(1)}(t)+ \tilde{E}_{n-2}^{(1)}(t)
\eeqa

where

\beq
\tilde{E}_{n}^{(1)}(t)=\int_o^t d\tau e^{-\beta_0 \tau/2}\sum_{ijk} E^{(0)}(t-\tau)
R_j E_k^{(0)}(\tau) \delta(n-i-j-k).
\eeq
The expressions of $E^{(0)}$ and $E^{(1)}$ are inserted into eq.~(\ref{solve}) thereby
providing a complete NLO solution of the singlet sector. The implementation of this method to NLO 
is quite involved and provides an accuracy which is comparable to other methods of solutions.

\section{Second view: the master form of the evolution equations and a hierarchy}

A second part of our brief review is dedicated to the discussion of the kinetic structure 
that underlines an evolution equation of DGLAP type.
 We recall that master equations are introduced 
in statistical mechanics in order to describe the evolution in time of some 
diffusive processes. 
One can take as an example an ensemble of random walkers moving on a given lattice. The density of 
walkers present in a certain point on the lattice can be described by a master equation and expressed in terms of transition probabilities. The master form of 
the DGLAP equation can be derived quite simply by performing some manipulations on the kernels, having separated the bulk contributions $(x<1)$ from the edge points $(x=1)$. Below we show how this is 
obtained and how one can contruct, by a formal Kramers-Moyal expansion of the transition probabilities, a hierarchy of PDE's of 
arbitrarily high orders.

Let's start considering  a generic 1-D master equation for transition
probabilities $w(x|x')$ which we interpret as the probability of making
a transition to a point $x$ given a starting point $x'$ for a given physical system.
The picture we have
in mind is that of a gas of particles making collisions in 1-D and entering the
interval $(x,x + dx)$ with a probability $w(x|x')$ per single transition,
or leaving it with a transition probability $w(x'|x)$.
In general one writes down a master equation
\beq
\frac{\partial }{\partial \tau}f(x,\tau)=\int dx'\left(
w(x|x') f(x',\tau) -w(x'|x) f(x,\tau)\right) dx'.
\eeq
describing the time $\tau$ evolution of the density of the gas
undergoing collisions or the motion of a many replicas of walkers of density
$f(x,\tau)$ jumping with a pre-assigned probability,
according to taste.

The result of Collins a Qiu, who were after a
derivation of the DGLAP equation that could include automatically also the ``edge point''
contributions (or x=1 terms of the DGLAP kernels) is in pointing out the existence of a
probabilistic picture of the DGLAP dynamics.
These edge point terms had been always introduced in the past only by hand and serve to enforce the baryon number sum rule and the momentum sum rule as $Q$, the momentum scale, varies.

\begin{figure}[t]
{\centering \resizebox*{6cm}{!}{\rotatebox{0}{\includegraphics{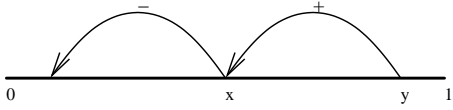}}} \par}
\caption{ The constrained random walk of the parton densities}
\label{walk}
\end{figure}

The kinetic interpretation was used in \cite{Teryaev} to provide an alternative proof
of Soffer's inequality.
We recall that this inequality
\beq
|h_1(x)| < q^+(x)
\eeq
famous by now, sets a bound on the transverse spin distribution $h_1(x)$ in terms of the
components of the positive helicity component of the quarks, for a given flavour.
The inequality has to be respected by the evolution.
We recall that $h_1$, also denoted by the symbol
\begin{equation}
\Delta _{T}q(x,Q^{2})\equiv q^{\uparrow }(x,Q^{2})-q^{\downarrow }(x,Q^{2}),
\end{equation}
has the property
of being purely non-singlet and of appearing at leading twist. It is
identifiable in transversely polarized
hadron-hadron collisions and not in Deep Inelastic Scattering (DIS), where can
appear only through an insertion of the electron mass in the unitarity
graph of DIS.

The connection between the Collins-Qiu form of the DGLAP equation and the 
master equation
is established as follows.
The DGLAP equation, in its original formulation is generically written as

\beq
\frac{d q(x,Q^2)}{d \log( Q^2)} = \int_x^1 \frac{dy}{y} P(x/y)q(y,Q^2),
\eeq
where we are assuming a scalar form of the equation, such as in the non-singlet sector. The generalization to the singlet sector of the arguments given below is, of course, quite straightforward.
To arrive at a probabilistic picture of the equation we start reinterpreting
$\tau=\log (Q^2)$ as a time variable, while the parton density $q(x,\tau)$
lives in a one dimensional (Bjorken) $x$ space.

We recall that the kernels are defined as ``plus'' distributions.
Conservation of baryon number, for instance, is enforced by the addition of
edge-point contributions proportional to $\delta(1-x)$.

We start with the following form of the kernel
\beq
P(z) = \hat{P}(z) - \delta(1-z) \int_0^1 \hat{P}(z)\, dz,
\label{form}
 \eeq
where we have separated  the edge point contributions from the rest
of the kernel, here called $\hat{P}(z)$. This manipulation is understood in all the
equations that follow.
The equation is rewritten in the following form

\beq
\frac{d}{d \tau}q(x,\tau) = \int_x^1 dy \hat{P}\left(\frac{x}{y}\right)\frac{q(y,\tau)}{y}
-\int_0^x \frac{dy}{y}\hat{ P}\left(\frac{y}{x}\right)\frac{q(x,\tau)}{x}
\label{bolz}
\eeq

Now, if we define
\beq
w(x|y)= \frac{\alpha_s}{2 \pi} \hat{P}(x/y)\frac{\theta(y > x)}{y}
\eeq
(\ref{bolz})
becomes a master equation for the probability function $q(x,\tau)$
\beq
\frac{\partial }{\partial \tau}q(x,\tau)=\int dx'\left(
w(x|x') q(x',\tau) -w(x'|x) q(x,\tau)\right) dx'.
\label{masterf}
\eeq
There are some interesting features of this special master equation. Differently from
other master equations, where transitions are allowed from  a given x both toward $y>x$
and $y< x$, in this case, transitions toward x take place only from values $y>x$ and
leave the momentum cell $(x, x+ dx)$ only toward smaller y values (see Fig.(\ref{walk}).

Clearly, this sets a direction of the kinetic evolution of the densities from large
x values toward smaller-x values as $\tau$, the fictitious ``time'' variable, increases.

Probably this is the simplest illustration of the fact that parton densities, at large final evolution scales, are dominated by their
small-x behaviour. As the ``randomly moving partons'' reach the $x\approx 0$ region
of momentum space, they find no space where to go, while other partons tend to pile up
toward the same region from above. This is the picture of a
random walk biased to move downward (toward the small-x region) and is illustrated in Fig.~(\ref{walk}).

\section{Probabilistic Kernels}

 We briefly discuss some salient features of the structure of the kernels in this approach and comment on the type
of regularization involved in order to define them appropriately.

We recall that unpolarized and polarized kernels, in leading
order, are given by
\beqa
P^{(0)}_{NS} &=&P_{qq}^{(0)}=C_F\left( \frac{2}{(1-x)_+} -1 - x +
\frac{3}{2} \delta(1-x)\right) \nonumber\\
P_{qg}^{(0)}&=& 2 T_f\left(x^2 + (1-x)^2)\right) \nonumber \\
P_{gq}^{(0)}&=& C_F \frac{1 + (1 -x)^2}{x}\nonumber \\
P_{gg}^{(0)}&=& 2 N_c\left(\frac{1}{(1-x)_+} +\frac{1}{x} -2 + x(1-x)\right)
 +\frac{\beta_0}{2}\delta(1-x)
\eeqa
where
\beq
C_F=\frac{N_C^2 -1}{2 N_C}, \,T_f=T_R n_f= \frac{1}{2} n_f,
\,\,\beta_0=\frac{11}{3} N_C  - \frac{4}{3} T_f \nonumber\\
\eeq
and
\beqa
\Delta P^{(0)}_{NS} &=&\Delta P_{qq}^{(0)}\nonumber\\
\Delta P_{qq}^{(0)}&=&C_F\left( \frac{2}{(1-x)_+} - 1 - x +
 \frac{3}{2}\delta(1-x)\right)\nonumber \\
\Delta P_{qg}^{(0)}&=& 2 T_f( 2 x - 1) \nonumber \\
\Delta P_{gq}^{(0)}&=& C_F (2 -x)\nonumber \\
\Delta P_{gg}^{(0)}&=& 2 N_c\left(\frac{1}{(1-x)_+} - 2 x + 1\right)
+\delta(1-x)\frac{\beta_0}{2},
\label{stand2}
\eeqa

while the LO transverse kernels are given by
\beq
\Delta_T P^{(0)}_{qq}=C_F\left( \frac{2 }{(1-x)_+} - 2  
+ \frac{3}{2}\delta(1-x)\right).
\eeq

The unpolarized kernels should be compared with the Collins-Qiu form 
\beqa
P_{qq}&=&\gamma_{qq} -\delta(1-x) \int_0^1 dz \gamma_{qq}
\nonumber \\
P_{gg}&=&\gamma_{gg} - 
\left(n_f \int_0^1 {dz} \gamma_{qg} +\frac{1}{2}\int_0^1 dz 
\gamma_{gg}\right) \delta(1-x) \nonumber \\
P_{qg}&=& \gamma_{qg}\nonumber \\
P_{gq}&=& \gamma_{gq}\nonumber \\
\label{cq1}
\eeqa
where
\beqa
\gamma_{qq}&=& C_F \left(\frac{2}{1-x} -1 - x 
\right)\nonumber \\
\gamma_{qg} &=& (2 x -1)\nonumber \\
\gamma_{gq} &=& C_F(2 - x) \nonumber \\
\gamma_{gg} &=& 2 N_c\left( \frac{1}{1-x} +\frac{1}{x} -2 + x(1-x) \right). \nonumber \\
\label{cq2}
\eeqa
These kernels need a suitable regularization to be well defined. 
Below we will analize the implicit regularization 
underlying eq.~(\ref{cq1}). One observation is however almost immediate: 
the component $P_{gg}$ is not of the form given by eq.~(\ref{form}). 
In general, therefore, in the singlet case, the generalization of 
eq.~(\ref{form}) is given by 
\beq
P(x)=\hat{P}_1(x) - \delta(1-x)\int_0^1 \hat{P}_2(z) dz
\eeq
and a probabilistic interpretation is more complex compared to the non-singlet 
case and has been discussed in the original literature \cite{CollinsQiu}. 

\section{Convolutions and Master Form of the Singlet}
Distributions are folded with the kernels and the result rearranged
in order to simplify the structure of the equations. Since
in the previous literature this is done in a rather involute way 
\cite{Hinchliffe} 
we provide here a simplificaton, from which the equivalence of the 
various forms of the kernel, in the various regularizations adopted, will be 
apparent.     
All we need is to apply (\ref{simplerel}) and the evolution equations become
\beqa
\frac{d q}{d \log(Q^2)} &=& 2 C_F 
\int\frac{dy}{y}\frac{ y q(y) - x q(x)}{y-x} +2 C_F \log(1-x)\, q(x) -
\int_x^1\frac{dy}{y}\left( 1 + z\right)q(y) + 
\frac{3}{2} C_F q(x) \nonumber \\
&& + n_f\int_x^1\frac{dy}{y}
\left( z^2 +(1-z)^2\right)g(y)\nonumber \\
\frac{d g}{d \log(Q^2)} &=& 
C_F \int_x^1\frac{dy}{y}\frac{1 +(1-z)^2}{z}q(y)
+ 2 N_c \int_x^1\frac{dy}{y}
\frac{ y f(y) - x f(x)}{y-x}g(y) 
\nonumber \\
&& + 2 N_c \log(1-x) g(x) 
+2 N_c\int_x^1 \frac{dy}{y}\left( \frac{1}{z} -2 + z(1-z)\right)g(y) + 
\frac{\beta_0}{2}g(x) \nonumber \\
\label{standard}
\eeqa
with $z\equiv x/y$. The same simplified form is obtained 
from the probabilistic version, having defined a suitable regularization 
of the edge point singularities in the integrals over the components 
$\gamma_{f f'}$ in eq. (\ref{cq2}). The canonical expressions of the 
kernels (\ref{stand2}), expressed in terms of ``+'' distributions, can also be rearranged to look like their equivalent probabilistic 
form by isolating the edge-point contributions hidden in their ``+''
distributions. We get the expressions

\beqa
{P_{qq}^{(0)}}_{NS} &=&P_{qq}^{(0)}=C_F\left( \frac{2}{(1-x)} -1 - x\right) - 
\left(C_F\int_0^1 \frac{dz}{1-z} -\frac{3}{2}\right) \delta(1-x) \nonumber\\
P_{gg}^{(0)}&=& 2 N_c\left(\frac{1}{(1-x)} +\frac{1}{x} -2 + x(1-x)\right)
-\left(2 N_c \int_0^1 \frac{dz}{1-z}-\frac{\beta_0}{2}\right)\delta(1-x) \nonumber \\
\eeqa

and
\beqa
\Delta P_{qq}^{(0)}&=&C_F\left( \frac{2}{(1-x)} - 1 - x\right) -
C_F\left( \int_0^1 \frac{dz}{1-z} - \frac{3}{2}\right)\delta(1-x)\nonumber \\
\Delta P_{gg}^{(0)}&=& 2 N_c\left(\frac{1}{1-x} - 2 x + 1\right)
-\left(2 N_c  \int_0^1 \frac{dz}{1-z}  -\frac{\beta_0}{2}\right) \delta(1-x),
\eeqa
the other expressions remaining invariant. 

A master form of the singlet (unpolarized) equation 
is obtained by a straightforward change of variable in the decreasing 
terms. We obtain 
\beqa
\frac{d q}{d \tau}&=&\int_x^{1 - \Lambda}\frac{dy}{y}\gamma_{qq}(x/y)q(y) 
-\int_0^{x - \Lambda}\frac{dy}{y}\gamma_{qq}(y/x) q(x)\nonumber \\
\frac{d g}{d \tau}&=&\int_x^{1 - \Lambda}\frac{dy}{y}\gamma_{gg}(x/y)
- n_f\int_0^x\gamma_{qg}(y/x) g(x)  \nonumber \\
&& -\frac{1}{2} \int_\Lambda^{x - \Lambda}\gamma_{gg}(y/x) g(x) + 
\int_x^1\frac{dy}{y}\gamma_{gq}(x/y)q(y)
\eeqa 
with a suitable (unique) cutoff $\Lambda$ needed to cast the 
equation in the form (\ref{standard}). 
The (regulated) transition probabilities are then given by 
\beqa
w_{qq}(x|y)&=&\gamma_{qq}(x/y)\frac{\theta(y>x)\theta(y< 1 -\Lambda)}{y}\nonumber \\  
w_{qq}(y|x)&=&\gamma_{qq}(y/x)\frac{\theta(y< x- \Lambda) \theta(y>0)}{x} \nonumber \\ 
w_{gg}(x|y) &=&\gamma_{gg}(x/y)\frac{\theta(y>x)\theta(y< 1 -\Lambda)}{y} \nonumber \\ 
w_{qq}(y|x)&=&\left(n_f\gamma_{qg}(y/x)
-\frac{1}{2}\gamma_{gg}(y/x)\right)\frac{\theta(y< x- \Lambda) \theta(y>0)}{x} \nonumber \\
w_{gq}(y|x)&=&\gamma_{gq}(x/y)\frac{\theta(y>x)\theta(y< 1 -\Lambda)}{y}\nonumber \\  
w_{gq}(x|y)&=& 0, \nonumber \\
\eeqa
as one can easily deduct from the form of eq. (\ref{masterf}).

\section{A Kramers-Moyal Expansion for the DGLAP Equation}
A way to get rid of the integrals that appear in the master equation is
the Kramers-Moyal (KM) expansion. These expansions (backward or forward)
are sometimes useful in order to gain insight into the master equation itself,
since they may provide a complementary view of the underlying dynamics.

The expansion introduces differential operator of arbitrary order, due to the nonlocal structure of 
the equation.  For the approximation to be useful, one has to stop the expansion after
the first few orders and in many cases this turns out to be possibile.
Examples of processes of this type are special Langevin processes and
processes described by a Fokker-Planck operator. In these cases
the probabilistic interpretation allows us to write down a fictitious
lagrangean, a corresponding path integral and solve for the
propagators using the Feynman-Kac fomula. 
 For definitess we take the integral to cover all
the real axis in the variable $x'$ 

\beq
\frac{\partial }{\partial \tau}q(x,\tau)=\int_{-\infty}^{\infty} dx'\left(
w(x|x') q(x',\tau) -w(x'|x) q(x,\tau)\right) dx'.
\eeq

As we will see below, in the DGLAP case some
modifications to the usual form of the KM expansion will appear. 
At this point we perform a KM expansion of the equation in the usual way.
We make the substitutions in the master equation $y\to x-y$ in the first term and
$y\to x + y$ in the second term

\beq
\frac{\partial }{\partial \tau}q(x,\tau)=\int_{-\infty}^{\infty} dy\left(
w(x|x -y) q(x - y,\tau) -w(x + y|x) q(x,\tau)\right) ,
\eeq
identically equal to
\beq
\frac{\partial }{\partial \tau}q(x,\tau)=\int_{-\infty}^{\infty} dy\left(
w(x + y - y'|x -y') q(x - y',\tau) - w(x + y'|x) q(x,\tau)\right),
\label{shift}
\eeq
with $y=y'$. First and second term in the equation above differ by a shift (in $-y'$) and can be related using a Taylor (or KM) expansion of the first term

\beqa
\frac{\partial }{\partial \tau}q(x,\tau)&=&\int_{-\infty}^{\infty} dy \sum_{n=1}^\infty
\frac{(-y)^n}{n!}\frac{\partial^n}{\partial x^n}\left( w(x + y|x)
q(x,\tau)\right)
\eeqa
where the $n=0$ term has canceled between the first and the second
contribution coming from (\ref{shift}). The result can be
written in the form
\beq
\frac{\partial }{\partial \tau}q(x,\tau)= \sum_{n=1}^\infty
\frac{(-y)^n}{n!}\frac{\partial^n}{\partial x^n}\left( a_n(x)q(x,\tau)\right)
\eeq
where
\beq
a_n(x)=\int_{-\infty}^{\infty} dy(y-x)^n w(y|x).
\eeq

In the DGLAP case we need to amend the former derivation, due to the
presence of boundaries $( 0 < x < 1)$ in the Bjorken variable $x$. For simplicity 
we will focus on the non-singlet case. 
We rewrite the master equation using the same change of variables used above 

\beqa
\frac{\partial}{\partial\tau}q(x,\tau) &=& \int_x^1 dy w(x|y)q(y,\tau) - 
\int_0^x dy w(y|x) q(x,\tau)
\nonumber \\
&& -\int_0^{\alpha(x)} dy w(x+y|x)* q(x,\tau)+ 
\int_0^{-x} dy w(x+y|x)q(x,\tau),
\eeqa

where the Moyal product gets simplified 
 
\beq
w(x+y|x)*q(x)\equiv w(x+y|x) e^{-y \left(\overleftarrow{\partial}_x + 
\overrightarrow{\partial}_x\right)} q(x,\tau)
\eeq

and $\alpha(x) =x-1$. 
The expansion is of the form 
\beqa
\frac{\partial}{\partial \tau}q(x,\tau)=\int_{\alpha(x)}^{-x}dy\,  w(x+y|x)q(x,\tau) - 
\sum_{n=1}^{\infty}\int_0^{\alpha(x)}dy \frac{(-y)^n}{n!}{\partial_x}^n
\left(w(x+y|x)q(x,\tau)\right)
\label{expans}
\eeqa
which can be reduced to a differential equation of arbitrary order using simple manipulations. 
We recall that the Fokker-Planck approximation is obtained stopping the expansion at 
the second order 
\beq
\frac{\partial}{\partial \tau}q(x,\tau)= a_0(x) -\partial_x\left(a_1(x) q(x)\right) + 
\frac{1}{2}\partial_x^2\left( a_2(x) q(x,\tau)\right)
\label{fpe}
\eeq
with 
\beq
a_n(x)=\int dy \,y^n\, w(x+y,x)
\eeq
being moments of the transition probability function $w$.
Given the boundary conditions on the Bjorken variable x, even in the 
Fokker-Planck approximation, the Fokker-Planck version of the DGLAP equation 
is slightly more involved than Eq. (\ref{fpe}) and the coefficients $a_n(x)$ need to be redefined.

\section{The Fokker-Planck Approximation}
Having identified a probabilistic description 
of the DGLAP equation in terms of a master equation, 
it is then natural to try to investigate the role of the Fokker-Planck (FP) approximation to it.
In the context of a random walk, an all-order derivative expansion of the master equation 
can be arrested to the first few terms either if the conditions of Pawula's 
theorem are satisfied -in which case the FP approximation turns out to be exact- 
or if the transition probabilities show an exponential decay above 
a certain distance allowed to the random walk. Since the DGLAP kernels 
show only an algebraic decay in x, 
and there isn't any explicit scale in the kernel themselves, 
the expansion is questionable. However, from a formal viewpoint, it is still allowed. 
With these caveats in mind we proceed to investigate the features of this expansion. 
  
We redefine 

\beqa
\tilde{a}_0(x) &=& \int_{\alpha(x)}^{-x} dy w(x+y|x)q(x,\tau) \nonumber \\
a_n(x) &=&\int_0^{\alpha(x)} dy y^n w(x+y|x) q(x,\tau) \nonumber \\
\tilde{a}_n(x)&=&\int_0^{\alpha(x)}
dy y^n {\partial_x}^n \left(w(x+y|x)q(x,\tau)\right) \,\,\,n=1,2,...
\eeqa
For the first two terms $(n=1,2)$ one can easily work out the relations
\beqa
\tilde{a}_1(x) &=&\partial_x a_1(x) - \alpha(x) \partial_x \alpha(x)
w(x + \alpha(x)|x)q(x,\tau) \nonumber \\
\tilde{a}_2(x) &=&\partial_x^2 a_2(x) - 
2 \alpha(x) (\partial_x \alpha(x))^2 w(x+ \alpha(x)|x) q(x,\tau) - 
\alpha(x)^2 \partial_x\alpha(x)
\partial_x\left( w(x+ \alpha(x)|x)q(x,\tau)\right)\nonumber \\ 
&& - \alpha^2(x)\partial_x \alpha(x) \partial_x\left( w(x + y|x)q(x,\tau)\right)|_{y=\alpha(x)}
\eeqa

Let's see what happens when we arrest the expansion (\ref{expans}) to the first 3 terms.
The Fokker-Planck version of the equation is obtained by including in the approximation
only $\tilde{a}_n$ with $n=0,1,2$.

The Fokker-Planck limit of the (non-singlet) equation is then given by
\beq
\frac{\partial}{\partial \tau}q(x,\tau) = \tilde{a}_0(x)
+\tilde{a}_1(x) - \frac{1}{2} \tilde{a}_2(x)
\eeq
which we rewrite explicitly as 
\beqa
\frac{\partial}{\partial \tau}q(x,\tau) &=& 
C_F\left(\frac{85}{12} +\frac{3}{4 x^4} - \frac{13}{3 x^3} +
\frac{10}{x^2} -\frac{12}{x}
+ 2\log\left(\frac{1-x}{x}\right)\right)q(x)\nonumber \\
&& +C_F\left( 9 - \frac{1}{2 x^3} +\frac{3}{x^2} -\frac{7}{x} 
-\frac{9}{2}\right) \partial_x q(x,\tau) \nonumber \\
&& + C_F\left(\frac{9}{4} +\frac{1}{8 x^2} -\frac{5}{6 x} -
\frac{5 x}{2}  +\frac{23 x^2}{24}\right) \partial_x^2 q(x,\tau).
\eeqa

A similar approach can be followed also for other cases, for which a 
probabilistic picture (a derivation of Collins-Qiu type) has not been established yet, 
such as for $h_1$. 
We describe briefly how to proceed in this case. 

First of all, we rewrite the evolution equation for the transversity 
in a suitable master form. This is possible since the subtraction terms 
can be written as integrals of a positive function. A possibility is 
to choose the transition probabilities
\beqa
w_1[x|y] &=& \frac{C_F}{y}\left(\frac{2}{1- x/y} - 2 \right)
\theta(y>x) \theta(y<1)\nonumber \\
w_2[y|x] &=& \frac{C_F}{x} \left(\frac{2}{1- y/x} - \frac{3}{2}\right)
\theta(y > -x)\theta(y<0)
\nonumber \\
\eeqa
which reproduce the evolution equation for $h_1$ in master form

\beq
\frac{d h_1}{d \tau}= \int_0^1 dy w_1(x|y)h_1(y,\tau) 
-\int_0^1 dy w_2(y|x) h_1(x,\tau).
\eeq
The Kramers-Moyal expansion is derived as before, with some slight
modifications. The result is obtained introducing an intermediate cutoff which is 
removed at the end. In this case we get 
\beqa
\frac{d h_1}{d\tau} &=& C_F\left( \frac{17}{3}
-\frac{2}{3 x^3} + \frac{3}{x^2} - \frac{6}{x}
+ 2 \log\left(\frac{1-x}{x}\right)\right) h_1(x,\tau) \nonumber \\
&& + C_F\left( 6 + \frac{2}{3 x^2} -\frac{3}{x} 
- \frac{11 x }{3}\right)\partial_x h_1(x,\tau) \nonumber \\
&& + C_F\left( \frac{3}{2} - \frac{1}{3 x } 
- 2 x + \frac{5 x^2}{6}\right)\partial_x^2 h_1(x,\tau).
\eeqa

Notice that compared to the standard Fokker-Planck approximation, the boundary now 
generates a term on the left-hand-side of the equation proportional to $q(x)$ 
which is absent in eq. (\ref{fpe}). This and higher order approximations to the 
DGLAP equation can be studied systematically both analytically and numerically 
and it is possible to assess the validity of the approximation \cite{CPC}.

\subsection{Links to Fractional Diffusion} 
A formal connection of the LO DGLAP equation to fractional diffusion can also be easily worked out 
\cite{CC} in dimensional regularization. Since 
the literature on fractional calculus and anomalous diffusion is quite vast \cite{Mainardi}, we need just few essential definitions to make our brief discussion self-contained. 
The n-th primitive of a distribution function $f(x)$ can be written as 

\beq
J^{n}f(x)=\frac{1}{(n -1)!}\int_x^1 (x-y)^{n-1} f(y) dy
\eeq
which can be easily analytically continued for any real $\alpha>0$
\beq
J^\alpha f(x)= \frac{1}{\Gamma[\alpha]}\int_x^1 f(y)(x-y)^{\alpha -1} dy  
\eeq
thereby defining the {\em fractional} integral of order $\alpha$ of the function $f(x)$.
We also recall that the {\em fractional derivative} of a function can be defined by a suitable analytic continuation
\beq
D^\beta f(x)= \frac{1}{\Gamma[n-\beta]}\frac{d^n}{dx^n} \int_x^1 \frac{f(y)}{(x-y)^{\beta-n+1}}dy 
\eeq
where  $n-1 <\beta < n$. In our case the role of $\beta$ is taken by the parameter $\epsilon$ of dimensional regularization. One can also formally 
define $D^{-\alpha}\equiv J^{\alpha}$ to denote the corresponding fractional primitive.

Now, since the kernels contain ``+'' distributions and Dirac delta functions, is is convenient to 
use the relation 

\beq
\left(\frac{1}{1- w}\right)^{1 + \epsilon}= \frac{1}{(1-w)_+} - \epsilon \left(\frac{\log(1-w)}{1-w}\right)_+ 
-\frac{1}{\epsilon}\delta(1-w) + O(\epsilon^2),
\eeq
valid in dimensional regularization, with $\epsilon >0$, from which one obtains 

\beq
\frac{1}{2}\left(\frac{1}{1- w}\right)^{1 + \epsilon}+\frac{1}{2} \left(\frac{1}{1- w}\right)^{1 - \epsilon} = \frac{1}{(1-w)_+}.
\eeq
 The equation for the transverse spin distribution, for instance, 
can be easily recast in the form 
\beq
\frac{\partial h_1}{\partial \tau} = - C_F \left(D^\epsilon + 
D^{-\epsilon}\right)h_1(x) -2 C_F D^{-1} \left(\frac{h_1(x)}{x}\right) + 
\frac{3}{2} C_F h_1(1) 
\eeq
with similar expressions for all the other non singlet equations. This equation is a possible 
starting point for the analysis of anomalous diffusion in the context of such equations. Equations of this type 
provide probability functions belonging to the class of stable distributions, such as those describing Levy processes  
and continuous time random walks. 
\section{Conclusions}

We have reviewed previous work of us on the determination of the solutions of renormalization group equations in 
QCD from x-space and on the KM expansion of the master formulation of these equations in analogy to the theory of 
stochastic processes.
Parton distributions represent the first important example of nonlocal operators in a realistic 
field theory that have been studied extensively for almost three decades. Although we are 
unable to compute from first principles these fundamental field theory structures, the study of their 
identification and classification at leading twist and higher has allowed to break substantial new ground in field theory and QCD phenomenology.


\begin{thebibliography}{10}
\bibitem{Sterman} G. Sterman, ``An introduction to Quantum Field Theory'', Cambridge Univ. Press, 1994;
J. C. Collins, D. E. Soper, G. Sterman, in 'Perturbative QCD' (A.H. Mueller, ed.) (World Scientific Publ., 1989), 
Adv.Ser.Direct.High Energy Phys.5, 1988. 
\bibitem{Gallipoli} A. Cafarella, C. Corianò, M. Guzzi 
Presented at International Workshop on Nonlinear Physics: Theory and Experiment, Gallipoli, 2002 (World Scientific)  
ed.  B. Prinari et. al., hep-ph/0209149.
\bibitem{zachos} T. Curtright and C. Zachos, Prog.Theor.Phys.Suppl. {\bf 135} 244, 1999;
J.Phys. {\bf A32} 771, 1999;  C. Zachos, Int.J.Mod.Phys. {\bf A17}, 2002.
\bibitem{RLJ} R.L. Jaffe, Nucl. Phys. B229 (1983) 205.
\bibitem{FP} W. Furmanski and R. Petronzio, Nucl.Phys.{\bf B195} 237, 1982.
\bibitem{CS} C. Corian\`o and C. Savkli, Comput.Phys.Commun.{\bf 118} 236,1999.
\bibitem{Cacorguz} A. Cafarella, C. Corian\`o and M. Guzzi, hep-ph/0512358.
\bibitem{FurmanskiPetronzio} W.~Furmanski and R.~Petronzio, Nucl.Phys.B \textbf{195} (1982) 237
\bibitem{CollinsQiu}J.C.~Collins and J.~Qiu, Phys.Rev.D \textbf{39} (1989) 1398
\bibitem{betafunction}T.~van Ritbergen, J.A.M.~Vermaseren and S.A.~Larin, Phys.Lett.B
\textbf{400} (1997) 379.
\bibitem{alpha_s}K.G.~Chetyrkin, B.A.~Kniehl and M.~Steinhauser, Phys.Rev.D. \textbf{79} (1997) 2184.
\bibitem{Rossi} G.~Rossi, Phys.Rev.D \textbf{29} (1984) 852
\bibitem{Storrow} J.H. Da Luz Vieira and J. K. Storrow,
Z.Phys. {\bf C51} (1991) 241.
\bibitem{Vogelsang} W. Vogelsang, Phys. Rev. {\bf D} \textbf{57} (1998), 1886.  
\bibitem{CPC} A.~Cafarella and C.~Corianò, Comput.Phys.Commun. \textbf{160} (2004)
213.
\bibitem{NNLO_nonsinglet}S.~Moch, J.A.M.~Vermaseren and A.~Vogt, Nucl.Phys.B \textbf{688}
(2004) 101.
\bibitem{NNLO_singlet}A.~Vogt, S.~Moch and J.A.M.~Vermaseren, Nucl.Phys.B \textbf{691}
(2004) 129.
\bibitem{Teryaev}C.~Bourrely, E.~Leader, O.V.~Teryaev, hep-ph/9803238
\bibitem{Hinchliffe}E.~Eichten, I.~Hinchliffe, K.D.~Lane and C.~Quigg, Rev.Mod.Phys.
\textbf{56} (1984) 579.
\bibitem{CafarellaCorianoGuzzi}A. Cafarella, C.Corian\`o and M.Guzzi, {\bf JHEP} 0311 059 (2003)
\bibitem{CafarellaCorianoGuzzi}A. Cafarella, C.Corian\`o and M.Guzzi, hep-ph/0303050
\bibitem{CC} C. Corian\`o, in progress.
\bibitem{Mainardi} R. Gorenflo and F. Mainardi, Lecture Notes of CISM, in 
``Fractals and Fractional Calculus in Continuum Mechanics'', Springer Verlag, 1997, 223. 
\end{thebibliography}
\end{document}